\documentclass[aps,showpacs]{revtex4}

\usepackage{graphicx,epsfig}
\begin{document}
\bibliographystyle{revtex}


\title{New physics and astronomy with the new gravitational-wave
observatories}

\author{Scott A.\ Hughes}
\email{hughes@itp.ucsb.edu}
\affiliation{Institute for Theoretical Physics, University of
California, Santa Barbara, CA 93106}
\author{Szabolcs M\'arka}
\email{smarka@ligo.caltech.edu}
\affiliation{LIGO Laboratory, California Institute of Technology,
Pasadena, CA 91125}
\author{Peter L.\ Bender}
\email{pbender@jila.colorado.edu}
\affiliation{JILA, University of Colorado and National Institute of
Standards and Technology, Boulder, Colorado 80309-0440}
\author{Craig J.\ Hogan}
\email{hogan@astro.washington.edu}
\affiliation{Astronomy and Physics Departments, University of
Washington, Seattle, Washington 98195-1580}

\date{15 October 2001}

\begin{abstract}
Gravitational-wave detectors with sensitivities sufficient to measure
the radiation from astrophysical sources are rapidly coming into
existence.  By the end of this decade, there will exist several
ground-based instruments in North America, Europe, and Japan, and the
joint American-European space-based antenna LISA should be either
approaching orbit or in final commissioning in preparation for launch.
The goal of these instruments will be to open the field of {\it
gravitational-wave astronomy}: using gravitational radiation as an
observational window on astrophysics and the universe.  In this
article, we summarize the current status of the various detectors
currently being developed, as well as future plans.  We also discuss
the scientific reach of these instruments, outlining what
gravitational-wave astronomy is likely to teach us about the universe.
\end{abstract}
\maketitle

\section{Introduction}

A common misconception outside of the gravitational-wave research
community is that the primary purpose of observatories such as LIGO is
to detect directly gravitational waves.  Although the first
unambiguous direct detection will certainly be a celebrated event, the
real excitement will come when gravitational-wave detection can be
used as an observational tool for astronomy.  Because the processes
which drive gravitational-wave emission are fundamentally different
from processes that radiate electromagnetically, gravitational-wave
astronomy will provide a view of the universe that is rather different
from our ``usual'' views.

Because they arise from fundamentally different physical processes,
the information carried by gravitational radiation is ``orthogonal''
to that carried by electromagnetic radiation.  Consider the following
differences:

\begin{itemize}

\item Electromagnetic waves are oscillations of electric and magnetic
fields that propagate through spacetime.  Gravitational waves are
oscillations in spacetime itself.

\item Astrophysical electromagnetic radiation typically arises from
the incoherent superposition of emissions from individual electrons,
atoms, and molecules.  They often provide direct information about the
thermodynamic state of a system or environment.  Gravitational waves
are coherent superpositions of radiation that arise from the bulk
dynamics of a dense source of mass-energy (matter or highly curved
spacetime).  They provide direct information about the system's
dynamics.

\item The wavelength of electromagnetic radiation is typically smaller
than the radiating system.  They can thus be used to form an image of
that system; any good public lecture on astronomy includes a number of
``pretty pictures''.  By contrast, the wavelength of gravitational
radiation is typically of order or larger than the size of the
radiating source.  Such waves {\it cannot} be used to image the
source; instead, the two gravitational-wave polarizations are akin to
sound, carrying a stereophonic description of the source's dynamics.
Indeed, it is becoming common for workers in gravitational radiation
to play audio encodings of expected gravitational-wave events, and the
notion of ``pretty sounds'' (or at least ``interesting sounds'') may
become widely accepted as gravitational-wave astronomy matures.

\item With a few exceptions, electromagnetic astronomy is based upon
deep imaging of narrow, small-angle fields of view: observers obtain a
large amount of information about sources on a small piece of the sky.
Gravitational-wave astronomy will be an all-sky affair: observatories
such as LIGO and LISA have nearly $4\pi$ steradian sensitivity to
events over the sky.  A consequence of this is that their ability to
localize a source's sky position is not good by usual astronomical
standards; but, it means that any source on the sky will be
detectable, not just sources towards which the instrument is
``pointed''.  The contrast between the all-sky sensitivity, relatively
poor angular resolution of gravitational-wave observatories, and the
pointed, high angular resolution of telescopes is very similar to the
angular resolution contrast of hearing and sight, strengthening the
useful analogy of gravitational waves with sound.

\item Electromagnetic waves interact strongly with matter;
gravitational-waves do not.  This is both blessing and curse: it means
that gravitational waves propagate from emission to Earth with
essentially zero absorption, making it possible to probe astrophysics
that is hidden or dark (for example, the coalescence and merger of
black holes; the collapse of a stellar core; the dynamics of the early
universe very soon after the big bang).  It also means that the waves
interact very weakly with gravitational-wave detectors, requiring
enormous experimental effort for assured detection.

\item The directly observable gravitational waveform $h$ is a quantity
that falls off as $1/r$.  Most electromagnetic observables are some
kind of energy flux, and so fall off with distance with a $1/r^2$ law.
Because of this, relatively small improvements in the sensitivity of
gravitational-wave detectors have an enormous impact on their ability
to make measurements: doubling a detector's sensitivity doubles the
distance to which sources can be detected, increasing the volume of
the universe in which sources are measurable by a factor of eight.
Every factor of two improvement in the sensitivity of a
gravitational-wave observatory increases the number of observable
sources by nearly an order of magnitude.

\end{itemize}

These differences motivate the efforts to inaugurate
gravitational-wave astronomy as a vital field.  In this article, we
will describe current efforts to develop ground-based detectors (Sec.\
{\ref{sec:groundbased}}, focusing in particular on the American LIGO
project) as well as the joint American-European space-based LISA
project (Sec.\ {\ref{sec:spacebased}}).  We also discuss in some
detail the science that can be expected from gravitational-wave
astronomy (Sec.\ {\ref{sec:sciencereach}}).  First, we briefly discuss
the properties of gravitational waves and their likely strength, and
sketch how a gravitational-wave detector is able to measure their tiny
effects.

\section{Gravitational waves and detectors: overview}
\label{sec:overview}

Gravitational radiation is a natural consequence of general relativity
(indeed, of any causal theory of gravitation; see {\cite{dank_thesis}}
for a fascinating discussion of early work by Laplace on gravitation
with a finite speed of propagation).  It was first described
more-or-less correctly by Albert Einstein in 1918 {\cite{bigal1}}.
(In an earlier work {\cite{bigal2}}, Einstein predicted gravitational
radiation, though with a rather incorrect description.)  General
relativity describes the radiation as a tensor perturbation to the
metric of spacetime (so that it can be thought of as a spin-2 boson),
propagating at the speed of light, and with two independent
polarizations.

As electromagnetic radiation is generated by the acceleration of
charges, gravitational radiation arises from the acceleration of
masses.  In particular, electromagnetic waves are created (at lowest
order) by the time changing charge dipole moment, and are thus dipole
waves.  Monopole EM radiation would violate charge conservation.  At
lowest order, gravitational waves come from the time changing
quadrupolar distribution of mass and energy; monopole gravitational
waves would violate mass-energy conservation, and dipole waves violate
momentum conservation.

\begin{figure}[t]
\includegraphics[width = 17cm]{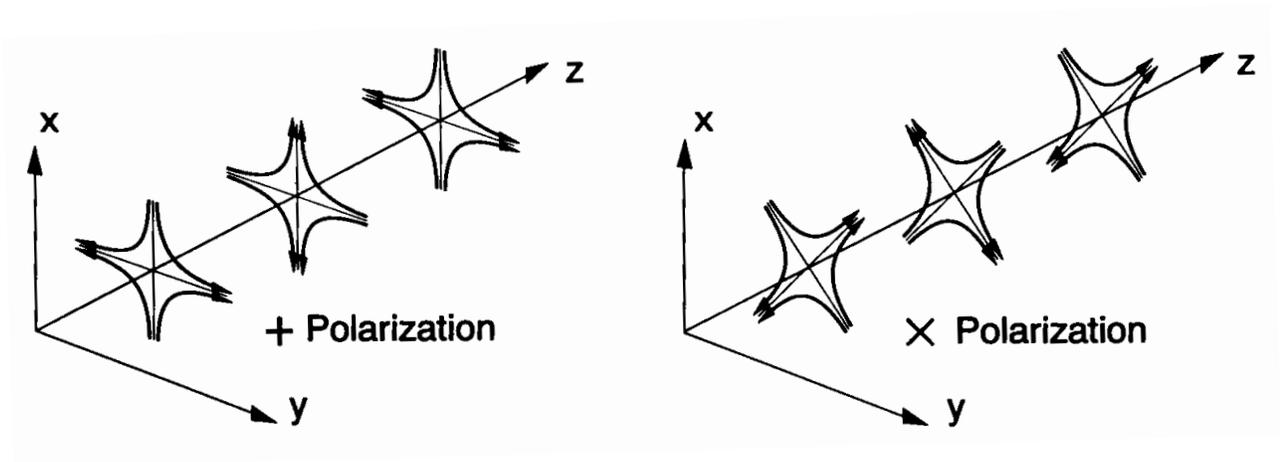}
\caption{The lines of force associated with the two polarizations of
a gravitational wave (from Ref.\ {\cite{ligoscience}}).}
\label{forcelines}
\end{figure}

Gravitational waves act tidally, stretching and squeezing any object
that they pass through.  Because the waves arise from quadrupolar
oscillations, they are themselves quadrupolar in character, squeezing
along one axis while stretching along the other.  When the size of the
object that the wave acts upon is small compared to the wavelength (as
is the case for LIGO), forces that arise from the two GW polarizations
act as in Fig.\ {\ref{forcelines}}.  The polarizations are named
``$+$'' (plus) and ``$\times$'' (cross) because of the orientation of
the axes associated with their force lines.

\begin{figure}[t]
\includegraphics[width = 17cm]{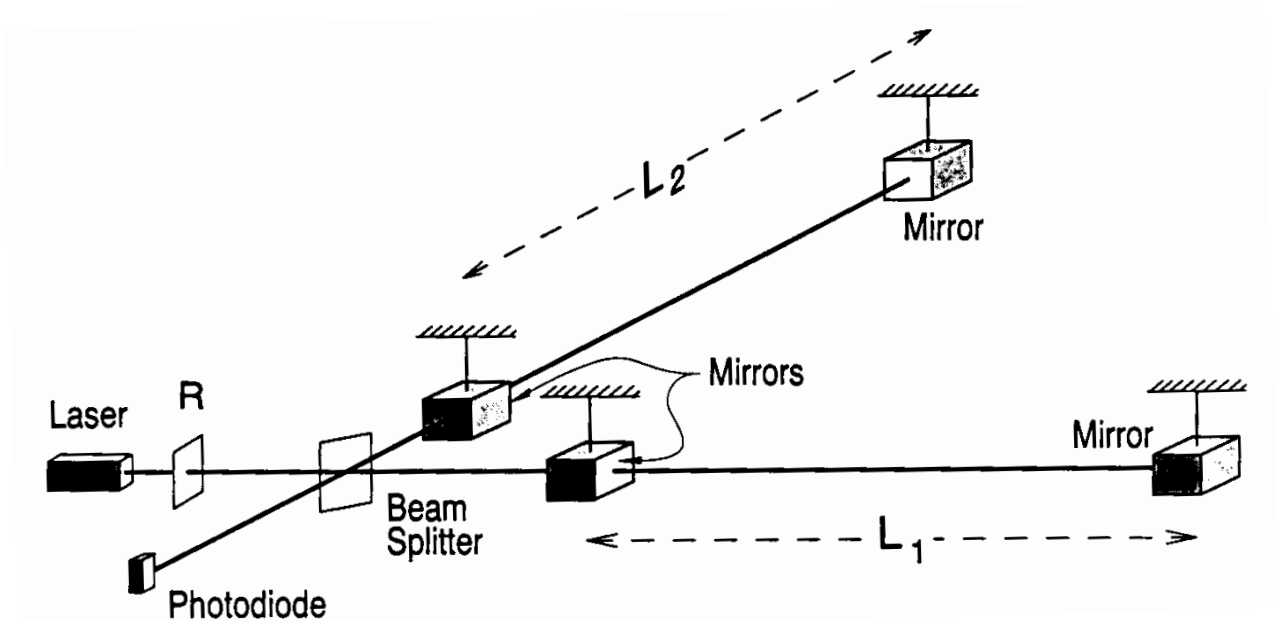}
\caption{Layout of an interferometer for detecting gravitational waves
(from Ref.\ {\cite{ligoscience}}).}
\label{fig:interferometer}
\end{figure}

\begin{figure}[t]
\includegraphics[width = 14cm]{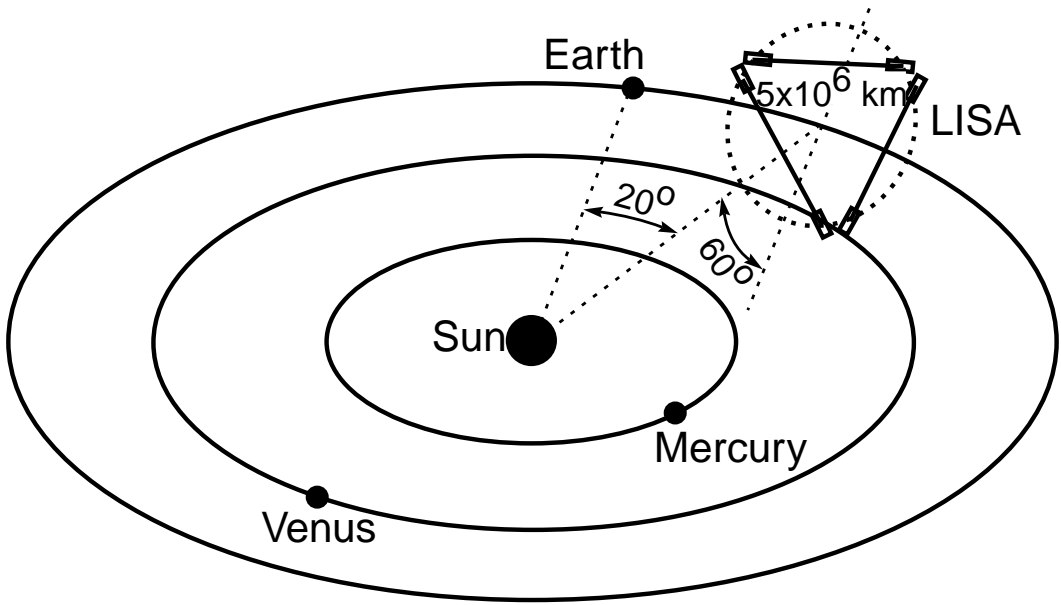}
\caption{Orbital configuration of the LISA antenna.}
\label{fig:lisa_orbit}
\end{figure}

Interferometric gravitational-wave detectors measure this tidal field
by measuring their action upon a widely-separated set of test masses.
In ground-based interferometers, these masses are arranged as in Fig.\
{\ref{fig:interferometer}}.  The space-based detector LISA arranges
its test masses in a large equilateral triangle that orbits the sun,
illustrated in Fig.\ {\ref{fig:lisa_orbit}}.  On the ground, each mass
is suspended with a sophisticated pendular isolation system to
eliminate the effect of local ground noise.  Above the resonant
frequency of the pendulum (typically of order $1\,{\rm Hz}$), the mass
moves freely.  (In space, the masses are actually free floating.)  In
the absence of a gravitational wave, the sides $L_1$ and $L_2$ shown
in Fig.\ {\ref{fig:interferometer}} are about the same length $L$.

Suppose the interferometer in Fig.\ {\ref{fig:interferometer}} is
arranged such that its arms lie along the $x$ and $y$ axes of Fig.\
{\ref{forcelines}}.  Suppose further that a wave impinges on the
detector down the $z$ axis, and the axes of the $+$ polarization are
aligned with the detector.  The tidal force of this wave will stretch
one arm while squeezing the other; each arm oscillates between stretch
and squeeze as the wave itself oscillates.  The wave is thus
detectable by measuring the separation between the test masses in each
arm and watching for this oscillation.  In particular, since one arm
is always stretched while the other is squeezed, we can monitor the
difference in length of the two arms:
\begin{equation}
\delta L(t) \equiv L_1(t) - L_2(t)\;.
\label{eq:delta_L}
\end{equation}
For the case discussed above, this change in length turns out to be
the length of the arm times the $+$ polarization amplitude:
\begin{equation}
\delta L(t) = h_+(t)L\;.
\label{eq:h_simple}
\end{equation}
The gravitational wave acts as a strain in the detector; $h$ is often
referred to as the ``wave strain''.  Note that it is a dimensionless
quantity.  Aficionados of general relativity can easily derive Eq.\
(\ref{eq:h_simple}) by applying the equation of geodesic deviation to
the separation of the test masses and using a gravitational-wave
tensor on a flat background spacetime to develop the curvature tensor;
see Ref.\ {\cite{300yrs}}, Sec.\ 9.2.2 for details.  We obviously do
not expect astrophysical gravitational-wave sources to align
themselves in as convenient a manner as described above.  Generally,
both polarizations of the wave influence the test masses:
\begin{equation}
{\delta L(t)\over L} = F^+ h_+(t) + F^\times h_\times(t) \equiv h(t)\;.
\label{eq:h_def}
\end{equation}
The antenna response functions $F^+$ and $F^\times$ weight the two
polarizations in a quadrupolar manner as a function of a source's
position and orientation relative to the detector; see
{\cite{300yrs}}, Eqs.\ (104a,b) and associated text.


In ground-based interferometers, the test masses at the ends of each
arm are made of a highly transparent material (fused silica in present
designs; perhaps sapphire in future upgrades).  The mirrors at the far
end of each arm have amplitude reflectivities approaching unity.  The
mirrors at the corner are less reflective since they must couple the
light into the Fabry-Perot cavity arms.  The corner mirrors'
multilayer dielectric coatings have power reflectivities $T \sim 3\%$.
A very stable laser beam is divided at the beamsplitter, directing
light into the two arm cavities.  If the finesse of the cavity is
${\cal F}$ and the amplitude reflectivity of the corner mirrors is
$r_{\rm corner}$, then each photon makes on average ${\cal
F}/\pi\simeq\sqrt{r_{\rm corner}}/(1 - r_{\rm corner}) \sim 65$
bounces.  The light from the two arms then recombines at the
beamsplitter.  The mirrors are positioned so that, in the absence of a
gravitational wave, all of the light goes back towards the laser and
the photodiode reads no signal.  If a signal is present, the relative
phase $\Phi$ of the two beams changes by an amount proportional to
$h$, changing the light's interference pattern.  Without any
intervention, this would cause light to leak into the photodiode.  In
principle, the wave strain $h$ could be read from the photointensity
of this light.  In practice, a system of servo loops controls the
system such that destructive interference is guaranteed --- the
photodiode is kept dark, and so is called the ``dark port''.  The wave
strain $h$ is then encoded in the servo signals used to keep the dark
port dark.

The wave strain must fall off with distance as a $1/r$ law to conserve
the total energy flowing through large spheres.  We have already
argued that the lowest order contribution to the waves is due to the
changing quadrupole moment of the source.  To order of magnitude, this
moment is given by $Q \sim (\mbox{source mass})(\mbox{source
size})^2$.  By dimensional analysis, we then know that the wave strain
must have the form
\begin{equation}
h \sim {G\over c^4}{\ddot Q\over r}\;.
\label{eq:h_ordermag}
\end{equation}
The second time derivative of the quadrupole moment is given
approximately by $\ddot Q\simeq 2M v^2\simeq 4 E^{\rm ns}_{\rm kin}$;
$v$ is the source's internal velocity, and $E^{\rm ns}_{\rm kin}$ is
the nonspherical part of its internal kinetic energy.  Strong sources
of gravitational radiation are sources that have strong non-spherical
dynamics --- for example, compact binaries (containing white dwarfs,
neutron stars, and black holes), mass motions in neutron stars and
collapsing stellar cores, the dynamics of the early universe.  In
order to have an interesting rate of observable events, we need to be
sensitive to sources at rather large distances.  For example, when
binary neutron stars coalesce, we need to reach out to several hundred
Mpc ({\it i.e.}, a substantial fraction of $10^9$ light years)
{\cite{nps,phinney91,kalogera_lorimer}}.  In such a case $E^{\rm
ns}_{\rm kin}/c^2 \sim\mbox{1 solar mass}\,(\equiv 1\,M_\odot)$.
Plugging these numbers into Eq.\ (\ref{eq:h_ordermag}) yields the
estimate
\begin{equation}
h \sim 10^{-21} - 10^{-22}\;.
\label{eq:h_num_est}
\end{equation}
This sets the sensitivity required to measure gravitational waves.
Combining this scale with Eq.\ (\ref{eq:h_def}) says that for every
kilometer of baseline $L$, we need to be able to measure a distance
shift $\delta L$ of better than $10^{-16}$ centimeters.

The prospect of achieving such a stringent displacement sensitivity
often strikes people as insane.  How can light whose wavelength
$\lambda\sim 10^{-4}\,{\rm cm}$ is $10^{12}$ times larger than the
typical displacement be used to measure that displacement?  For that
matter, how is it possible that thermal motions do not wash out such a
tiny effect?

That such measurement is possible with laser interferometry was
analyzed thoroughly and published by Rainer Weiss in 1972
{\cite{weiss72}}.  (It should be noted that the possibility of
detecting gravitational waves with laser interferometers has an even
longer history, reaching back to Pirani in 1956, and has been
independently invented by several workers: Gertsenshtein and Pustovoit
in 1962, Weber in the 1960s, and Weiss c.\ 1970.  See Sec.\ 9.5.3 of
Ref.\ {\cite{300yrs}} for further discussion and references.)  Examine
first how the 1 micron laser can measure a $10^{-16}$ cm effect.  As
mentioned above, the light bounces roughly 100 times before leaving
the arm cavity (corresponding to about half a cycle of a 100 Hz
gravitational wave).  The light's acquired phase shift during those
100 round trips is
\begin{equation}
\Delta\Phi_{\rm GW}\sim100\times2\times\Delta L\times2\pi/\lambda
\sim 10^{-9}\;.
\label{eq:phase_estimate}
\end{equation}
This phase shift can be measured provided that the photon shot noise
at the photodiode, $\Delta\Phi_{\rm shot}\sim 1/\sqrt{N}$, is less
than $\Delta\Phi_{\rm GW}$.  $N$ is the number of photons accumulated
over the measurement; $1/\sqrt{N}$ is the magnitude of phase
fluctuation in a coherent state, appropriate for describing a laser.
We therefore must accumulate $10^{18}$ photons over the roughly $0.01$
second measurement, which translates to a laser power of about 100
watts.  In fact, as was pointed out by Ron Drever
{\cite{drever_recycle}}, one can use a much less powerful laser: even
in the presence of a gravitational wave, only a tiny portion of the
light that comes out of the interferometer's arms goes to the
photodiode.  The vast majority of the laser power is sent back to the
laser.  An appropriately placed mirror bounces this light back into
the arms, {\it recycling} the laser light.  The recycling mirror is
shown in Fig.\ {\ref{fig:interferometer}}, labeled ``R''.  With it, a
laser of $\sim 10$ watts drives several hundred watts to circulate in
the ``recycling cavity'' (the optical cavity between the recycling
mirror and the arms), and $\sim 10$ kilowatts to circulate in the
arms.

Thermal excitations are overcome by averaging over many many
vibrations.  For example, the atoms on the surface of the
interferometers' test mass mirrors oscillate with an amplitude
\begin{equation}
\delta l_{\rm atom} = \sqrt{k T\over m\omega^2}
\sim 10^{-10}\,{\rm cm}
\end{equation}
at room temperature $T$, with $m$ the atomic mass, and with a
vibrational frequency $\omega\sim10^{14}\,{\rm s}^{-1}$.  This
amplitude is huge relative to the effect of the gravitational wave ---
how can we possibly hope to measure the wave?  The answer is that
atomic vibrations are random and incoherent.  The $\sim 7$ cm wide
laser beam averages over about $10^{17}$ atoms and at least $10^{11}$
vibrations in a typical measurement.  Atomic vibrations are irrelevant
compared to the coherent effect of a gravitational wave.  Other
thermal vibrations, however, are not irrelevant and in fact dominate
the noise spectrum of LIGO in certain frequency bands.  For example,
the test masses' normal modes are thermally excited.  The typical
frequency of these modes is $\omega\sim 10^5\,{\rm s}^{-1}$, and they
have mass $m \sim 10\,{\rm kg}$, so $\delta l_{\rm mass} \sim
10^{-14}\,{\rm cm}$.  This, again, is much larger than the effect we
wish to observe.  However, the modes are very high frequency, and so
can be averaged away provided the test mass is made from material with
a very high quality factor $Q$.  Understanding the physical nature of
noise in gravitational-wave detectors is an active field of current
research; see Refs.\
{\cite{levin,liu_thorne,santamore_levin,buonanno_chen,hughes_thorne,tcreighton}}
and references therein for a glimpse of recent work.  In all cases,
the fundamental fact to keep in mind is that a gravitational wave acts
{\it coherently}, whereas noise acts {\it incoherently}, and thus can
be beaten provided one is able to average away the incoherent noise
sources.

\section{Ground-based detectors}
\label{sec:groundbased}

The first generation of long baseline, kilometer-scale interferometric
gravitational-wave detectors are being constructed and commissioned at
several sites around the world.  Briefly, the major ground-based
interferometric gravitational-wave projects are as follows:

\begin{itemize}
\item \textbf{LIGO}.
Three LIGO interferometers are in the commissioning phase: two in
Hanford, Washington (with 2 and 4 km arms, sharing the vacuum system),
and one in Livingston, Louisiana (4 km arms).  An aerial view of the
Hanford site is included in Fig.\ {\ref{fig:ligo_optical}}.  The LIGO
detectors are designed to operate in the power recycled Michelson
configuration with arms acting as Fabry-Perot cavities.  The large
distance between sites (about $3000$ km) and differing arm lengths are
designed to support coincidence analysis.  Much current research and
development is focused on Advanced LIGO detector design.  The new
generation of detectors will provide a broader frequency band and a
$\sim$ 10-fold increase in range for inspiral sources via the lowered
noise floor.

\item \textbf{Virgo}.
Virgo, the Italian/French long baseline gravitational-wave detector,
is under construction near Pisa, Italy {\cite{marion2000}}.  It has 3
km arms and advanced passive seismic isolation systems.  In most
respects, Virgo is very similar to LIGO; a major difference is that it
should achieve better low frequency sensitivity from the beginning due
to its advanced seismic isolation system.  Virgo will very usefully
complement the LIGO detectors, strengthening coincidence analysis and
making source position determination possible.

\item \textbf{GEO600}.
GEO600 is a 600 meter interferometer being built by a British-German
collaboration in the vicinity of Hannover, Germany {\cite{luck2000}}.
It will use advanced interferometry and advanced low noise multiple
pendulum suspensions, serving as a testbed for advanced detector
technology and allowing it to achieve sensitivity comparable to the
multi-kilometer instruments.

\item \textbf{TAMA300}.
The TAMA detector near Tokyo, Japan can already claim significant
observation time with more than 1000 hours of operation
{\cite{andoetal}}.  It has achieved a peak strain sensitivity of
$h\sim 10^{-20}\,{\rm Hz}^{-1/2}$ {\cite{explain_sensitivity}} at
frequencies near 1000 Hz.  TAMA has 300 meter arms and is operated in
the recombined Michelson configuration with Fabry-Perot arms.  A much
improved 3 km detector is currently under design {\cite{kuroda2000}}.

\item \textbf{ACIGA}.
The Australian Consortium for Interferometric Gravitational Astronomy
plans to build an observatory near Perth, Australia
{\cite{mccleland2000}}.  They are presently engaged in the
construction of an 80 meter research interferometer, which can be
extended to kilometer scale.  They are studying advanced detection
methods and technologies which could lead to much decreased noise
floors in advanced ground based interferometers.

\end{itemize}

Since interferometric gravitational-wave detectors are sensitive to
nearly all directions, it is nearly impossible to deduce pointing
information from the signal of a single detector.  To get accurate
information about the source direction it is necessary to make use of
the detection time difference among detectors.  These detectors must
be widely spaced and not collinear.  At the very minimum, three sites
are needed for acceptable pointing.  A fourth detector, widely removed
from the plane of the other three, is particularly valuable for
improving directional information.  Thus a detector in Australia would
greatly add to the science output of the gravitational-wave
observatory network, which is otherwise confined entirely to the
northern hemisphere.

LIGO is a good example to illustrate the design and operation
principles of kilometer-scale interferometers; we shall focus on it
for the remainder of our discussion.

\subsection{LIGO overview}

Construction of both LIGO facilities and vacuum systems was completed
by early 2000.  The 2 km interferometer at the Hanford site was
installed in mid 2000; the Livingston site 4 km interferometer was
finished by late 2000.  Installation of the Hanford 4 km
interferometer was delayed due to repairs needed following the
February 28, 2001 earthquake; as this document is being written,
installation is well underway.  Both observatories are concentrating
on detector commissioning and a series of ``Engineering Runs''.  These
Runs have provided excellent real life experience for the
participating members of the LIGO Scientific Community, paving the way
for the approaching ``Science Run'' scheduled to begin at the end of
2001.  This will be followed by more frequent and longer Science Runs
until 2006, when major detector upgrades are scheduled.  In parallel
to the commissioning effort and Engineering/Science Runs, the
collaboration also focuses on the research and development of advanced
LIGO detectors, promising a wider detector band and greatly improved
sensitivity.

\begin{figure}[t]
\includegraphics[width = 16cm]{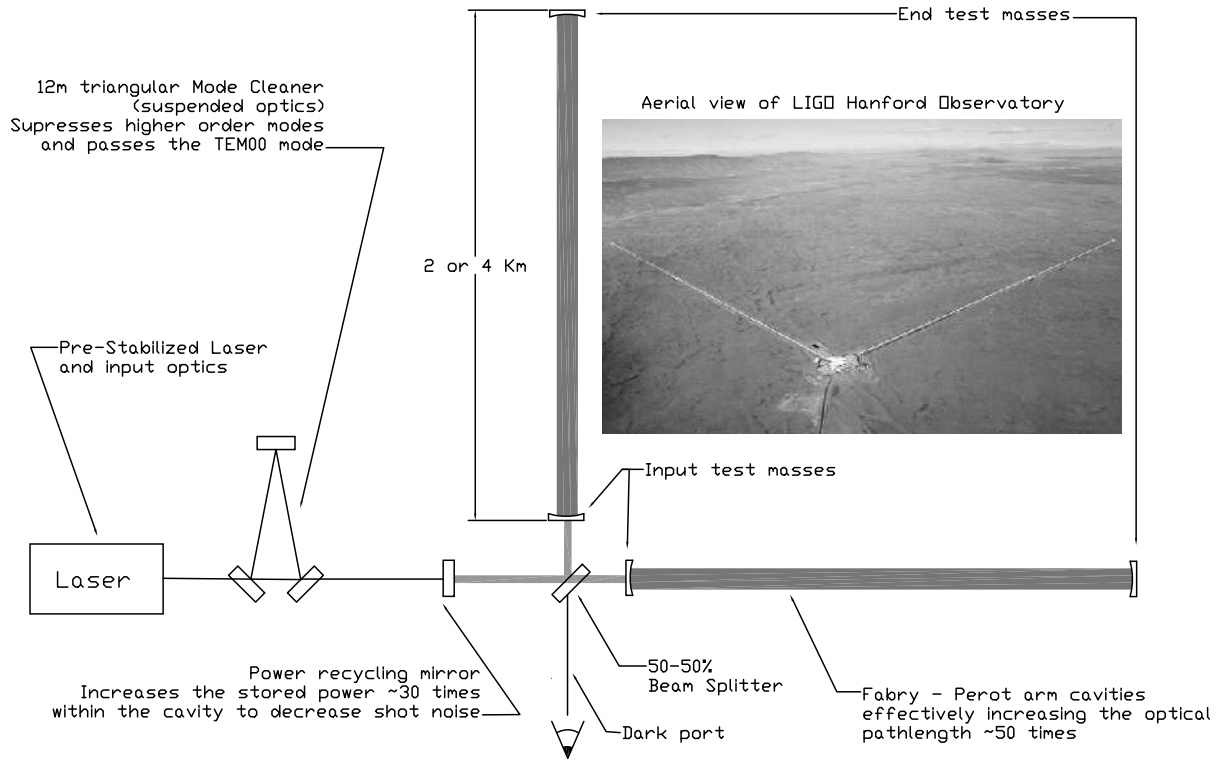}
\caption{Simplified optical layout of a LIGO interferometer.  Shown
here are the prestabilized laser, the input mode cleaner, the
recycling mirror, and the test mass mirrors.  As discussed in Sec.\
{\ref{sec:overview}}, servo loops ensure that the recombined light
destructively interferes so that the dark port is kept dark.  The
gravitational wave signal is read out from the forces needed to keep
the recombined light in destructive interference.}
\label{fig:ligo_optical}
\end{figure}

The LIGO detectors operate as power recycled Michelson interferometers
with Fabry-Perot arms; see Fig.\ {\ref{fig:ligo_optical}}.  Very high
duty cycle is needed for each interferometer in order to effectively
use the full network for coincidence analysis, which is necessary to
achieve a low false detection rate and have high confidence in
observations.  The wide (3000 km) separation between the LIGO sites is
large enough that the chance of environmentally induced coincidence
events is negligible.  Both sites are heavily equipped with
environmental sensors that thoroughly cover a wide range of possible
disturbances that could cause false detections.  For example, LIGO
monitors the local seismic background, electromagnetic fluctuations,
acoustic noise, cosmic radiation, dust, vacuum status, weather,
records power line transients, and uses ultra-sensitive magnetometers
at several locations at each observatory.

We now briefly describe the operating principles of LIGO optics, the
major sources of noise that limit sensitivity, the data analysis
system that will be used during operations, and plans for future
upgrades.

\subsubsection {Laser, optics and configuration}

The basic optical layout of the LIGO detectors is shown in Figure
{\ref{fig:ligo_optical}}.  LIGO uses a Nd:YAG near infrared laser
(wavelength 1064 nm) with peak power $\sim$ 10W as the light source.
Various electro-optical components and servo loops are used to
stabilize both the frequency and power of the laser.

The light from the pre-stabilized laser passes through the input
optics and is coupled into the 12 meter, triangular mode cleaner
cavity.  The mode cleaner passes only the TEM$_{00}$ mode, eliminating
higher order modes.  Starting with the mode cleaner, every major
optical component is within a large vacuum system, capable of reaching
$10^{-9}$ Torr.  After conditioning by the mode cleaner, the light
enters the interferometer.  All major optics in the interferometer are
suspended on a single steel wire loop, mechanically isolated from the
ground by vibration isolators and controlled by multiple servo loops.
The mirrors are made of fused silica with extremely high mechanical
$Q$ and are polished to within $\sim$ 1 nanometer RMS.  They have high
homogeneity, low bulk loss and multi-layer coatings with less than 50
ppm scattering loss.  Each mirror is actuated by four precision coils,
each positioned around a permanent magnet glued to the back side of
the test masses.  The coil assembly also features a sensitive shadow
sensor for local control.  Additional optical levers and wavefront
sensors provide more precise sensing.  The laser beam is coupled into
the arms by a beam splitter.  Each arm is a Fabry-Perot optical
cavity, increasing the effective length of the arm to magnify the
phase shift (proportional to cavity finesse) caused by the wave.  The
stored power within the interferometer is built up by the partially
transmitting recycling mirror.

An operating interferometer tries to keep the dark port perfectly
dark, holding the various optical components such that light coming
out of the arms destructive interferes and no light goes to the
photodetector.  When this is achieved, the interferometer is on {\it
resonance}, with maximum power circulating in the arms.  Several
interconnected control loops are used to achieve and then maintain
resonance.  An interferometer on resonance is usually described as
{\it locked}.  Keeping lock must be highly automated, requiring
minimal operator interaction for high uptime.  The gravitational-wave
signal is extracted from the servo signals used to maintain the lock
and correct for the length difference between the arms.

\subsubsection {Noise sources}

Gravitational-wave interferometers have their sensitivity limited by a
number of noise sources.  We list here some of the most important and
interesting fundamental noise sources; many of these were recognized
and had their magnitude estimated by Rainer Weiss in 1972
{\cite{weiss72}}.

\begin{itemize}

\item \textbf{Seismic noise}.
Ambient or culturally induced seismic waves continuously pass under
the test masses of the detector. The natural motion of the surface
peaks around 150 mHz; this is called the ``microseismic peak''.
Cultural noises tend to be at higher frequencies, near several Hz.
The test masses must be carefully isolated from the ground to
effectively mitigate the seismic noise.  Seismic noise will limit the
low frequency sensitivity of first generation ground-based
gravitational-wave detectors.

\item \textbf{Thermal noise}.
Thermally excited vibrational modes of the test mass or the suspension
system will couple into the resonances of the system.  By improving
the $Q$ of the components one can decrease the thermally induced noise
between resonances.

\item \textbf{Shot noise}.
The number of photons in the input laser beam fluctuates; this
surfaces as noise at the dark port.  The strain noise due
to this effect is proportional to $1/\sqrt{\mbox{(recycling
gain)(input laser power)}}$.  Thus, increasing the recycling gain
and/or increasing the laser power lowers the shot noise.
Unfortunately, high power in the cavities induces other unwanted
effects such as radiation pressure noise (discussed in the next item)
or thermal lensing (local deformation of the optical surfaces of the
cavity).  The right choice of laser power and recycling gain is a
compromise.

\item \textbf{Radiation pressure noise}.
Fluctuating number of photons bouncing from the mirrors will introduce
a fluctuating force on the mirror.  This effect is proportional to
$\sqrt{\mbox{(recycling gain)(input laser power)}}$ --- the inverse of
the proportionality entering the shot noise.  One cannot just increase
the laser power without penalty.  Reducing shot noise and radiation
pressure noise in tandem is a topic of advanced detector R\&D;
techniques for doing so require specially prepared laser states.  See
{\cite{buonanno_chen}} and references therein for further discussion.

\item \textbf{Gravity gradient noise}.
When seismic waves, atmospheric pressure fluctuations, cars, animals,
tumbleweeds, etc., pass near a gravitational-wave detector, they act
as density perturbations in the neighboring region.  This in turn can
produce significant fluctuating gravitational forces on the
interferometer's test masses {\cite{hughes_thorne,tcreighton}}.  This
is expected to become the limiting factor at low frequencies for
advanced ground-based detectors with high quality seismic isolation
systems.

\item \textbf{Laser intensity and frequency noise}.
The laser itself inevitably is somewhat noisy, with fluctuations in
both intensity and frequency.  This noise will not cancel out
perfectly when the signal from the two arms destructively recombines,
and so some noise can propagate into the signal on the dark port.

\item \textbf{Scattered light}.
Some laser light can scatter out of the main beam, and then be
scattered back, coupling into the interferometer's signal.  This light
will carry information about its scattering surface and be out of
phase with the beam, and thus contaminate the desired signal.  A dense
baffling system has been installed to greatly reduce this source of
noise.

\item \textbf{Residual gas}.
Any vacuum system contains some trace amount of gas that is extremely
difficult to reduce; in LIGO, these traces (mostly hydrogen) are at
roughly $10^{-9}$ Torr.  Density fluctuations from these traces in the
beam path will induce index of refraction fluctuations in the arms.
Residual gas particles bouncing off the mirrors can also increase the
displacement noise.

\item \textbf{Beam jitter}.
Jitter in the optics will cause the beam position and angle to
fluctuate slightly, which causes some noise at the dark port.

\item \textbf{Electric fields}.
Fluctuations in the electric field around the test masses can couple
into the interferometric signal via interaction between the field and
the induced or parasitic surface charge on the mirror surface.

\item \textbf{Magnetic fields}.
Fluctuations in the local magnetic field can affect the test masses
when interacting with the actuator magnets bonded to the surface of
the mirrors.

\item \textbf{Cosmic showers}.
High energy penetrating muons can be stopped by the test masses and
induce a random transient due to the recoil.

\end{itemize}

\begin{figure}[t]
\includegraphics[width = 12cm]{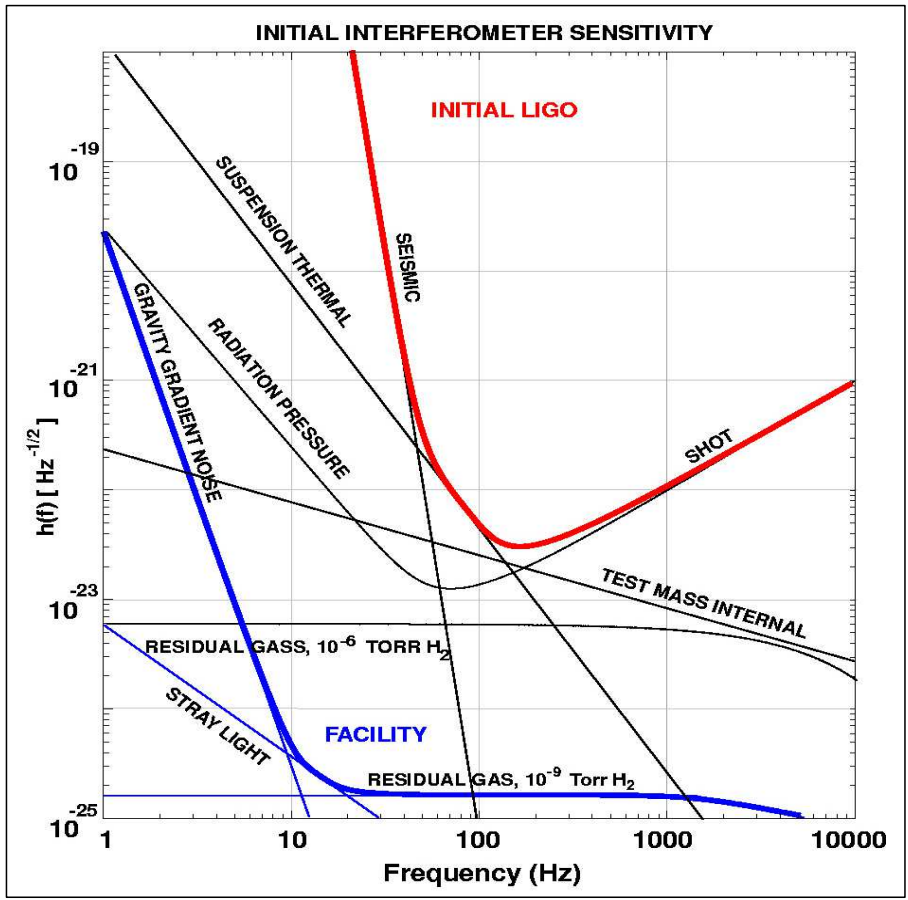}
\caption{Illustration of LIGO sensitivity, showing the amplitude
strain noise spectrum for various noise sources.  The heavy line
labeled ``Initial LIGO'' is the design goal for initial LIGO
interferometers; that labeled ``Facility'' shows the facility
limitations that cannot be avoided even if other noise sources are
perfectly controlled.}
\label{fig:ligo_sens}
\end{figure}

The impact of the most important of these noise sources is shown in
Fig.\ {\ref{fig:ligo_sens}}.  The initial LIGO detectors will be
limited by seismic noise at low frequencies ($\lesssim 50\,{\rm Hz}$),
by thermal noise in the test mass suspensions at intermediate
frequencies ($\sim 50 - 200\,{\rm Hz}$, and by shot noise at high
frequencies ($\gtrsim 200\,{\rm Hz}$).  The present detector noise is
well above this target, particularly at low and intermediate
frequencies.

As these noise sources are reduced, new challenges emerge.
Eventually, any given instrument will be limited by a set of noise
floors.  The curve labeled ``Facility Limit'' in Fig.\
{\ref{fig:ligo_sens}} sets the ultimate sensitivity that is likely to
be attained even with all other noise sources controlled perfectly.
At low frequencies, gravity gradient noise (which will be extremely
difficult, if not impossible, to mitigate) will be the dominant source
of noise; beyond that, optical path fluctuations from residual gas
limit the sensitivity.

\subsubsection {LIGO Data Analysis System (LDAS)}

As will be discussed further below, the gravitational-wave datastream
will be filtered by a large number ($\sim 10^4 - 10^5$) of model
waveforms (``templates'') to search for sources such as coalescing
compact binaries.  These templates must be applied in near real time.
Besides the gravitational-wave signal, LIGO detectors have a large
number of data channels for auxiliary data sources like the
environmental monitors discussed above.  A significant portion of
these signals must be digitized and recorded.  LIGO thus creates a
huge amount of data ($\sim 10$ MB per second, gathered around the
clock, equivalent to about 0.25 terabytes per day per detector) which
must be stored and managed in near real time.

To this end, LIGO has developed its own, state-of-the-art data
management system called LDAS (LIGO Data Analysis System).  LDAS is a
distributed computing environment, mixing remote process control on
servers with message passing on distributed clusters.  It provides a
framework for conducting scientific studies of LIGO data, allowing
data access and conditioning, signal reconstruction, coincidence
analysis, database management, data caching and archiving.  Large
Beowulf-type clusters will be used for analysis at LIGO's sites and at
the campuses.

The digitized data (at rates from 16 to 16384 samples per second) are
organized and stored in ``frames'', an international standard format
for gravitational-wave observatories.  The data are buffered on local
disks and archived on tape in the observatories.  The data include
accurate (sub microsecond) timestamps to facilitate coincidence
analyses between various different gravitational-wave detectors, and
also with other astrophysics experiments, such as neutrino or
gamma-ray detectors.  Parallel to LDAS, the data is digested by the
Global Diagnostic System, which is responsible for the real time
monitoring of environmental problems and the interferometer's state of
state.  All important local or externally generated events, from lock
losses to gamma ray bursts, are recorded in the LDAS metadata database
tables.  These records are available to analysis processes running
under LDAS.

\subsubsection{Detector upgrades}

A major focus of research within the gravitational-wave experimental
community right now is on developing technologies for improving the
sensitivity of LIGO and other ground-based detectors.  The first stage
detectors being commissioned right now are somewhat conservatively
designed, ensuring that they can be operated without needing to
develop too much new technology.  The price for this conservative
design is limited astrophysical reach: their sensitivity is such that
detection of sources is plausible, but not necessarily probable, based
on our current understanding of sources.

To broaden the astrophysical reach of these instruments, major
upgrades are planned for roughly the year 2006.  These upgrades will
push the lower frequency ``wall'' to lower frequencies, and push the
noise level down by a factor $\sim 10$ across the band.  This will
increase the distance to which sources can be detected by a factor of
$10 - 15$, and the volume of the universe which LIGO samples by a
factor of $1000 - 3000$.  This will dramatically boost the rate at
which events are measured.

Discussion of how these plans will be implemented is given in Ref.\
{\cite{gdsw}}.  Major changes will include a redesigned seismic
isolation system (pushing the wall down to about 10 Hz), a more
powerful laser (pushing the shot noise down by about a factor of 10),
and replacement of the optical and suspension components with improved
materials to reduce the impact of thermal noise.  In addition, the
system will allow ``tunable'' noise curves --- experimenters will be
able to shape the noise curve to some extent to chase after
particularly interesting sources.

\subsection{Sources for ground-based detectors}

Ground-based gravitational-wave detectors, particularly in their
earliest generations, are primarily sensitive only to extremely
violent astrophysical processes.  In the frequency band of interest
(roughly $100 - 1000$ Hz for early designs, and roughly $10 - 1000$ Hz
in future upgrades), these processes include the coalescence of
compact binary systems and stellar core collapse in supernova.  These
are extremely energetic but short-lived events; for example, near the
end of binary black hole coalescence, the binary's gravitational-wave
luminosity approaches the theoretical maximum $L \sim c^5/G \sim
10^{59}$ erg/sec for $\mbox{several }\times(10^{-2}-10^{-3})$ seconds,
brighter than any other other source in the sky.  Sources that should
be of interest in later generations are continuous gravitational-wave
emitters, such as pulsars or accreting neutron stars, and stochastic
backgrounds, perhaps relics left from the big bang.  Joint analyses
with data from neutrino and gamma-ray detectors may prove particularly
valuable, providing views of events through multiple radiation
channels.

In the remainder of these section, we discuss certain reasonably
well-understood sources of gravitational waves, focusing in particular
on those that promise to be useful in providing new tests of physics.
It is worth emphasizing at this point that, due to the pioneering
nature of this field, there is great hope that we will see significant
signals from unknown or unexpected sources.

\begin{figure}[t]
\includegraphics[width = 15cm]{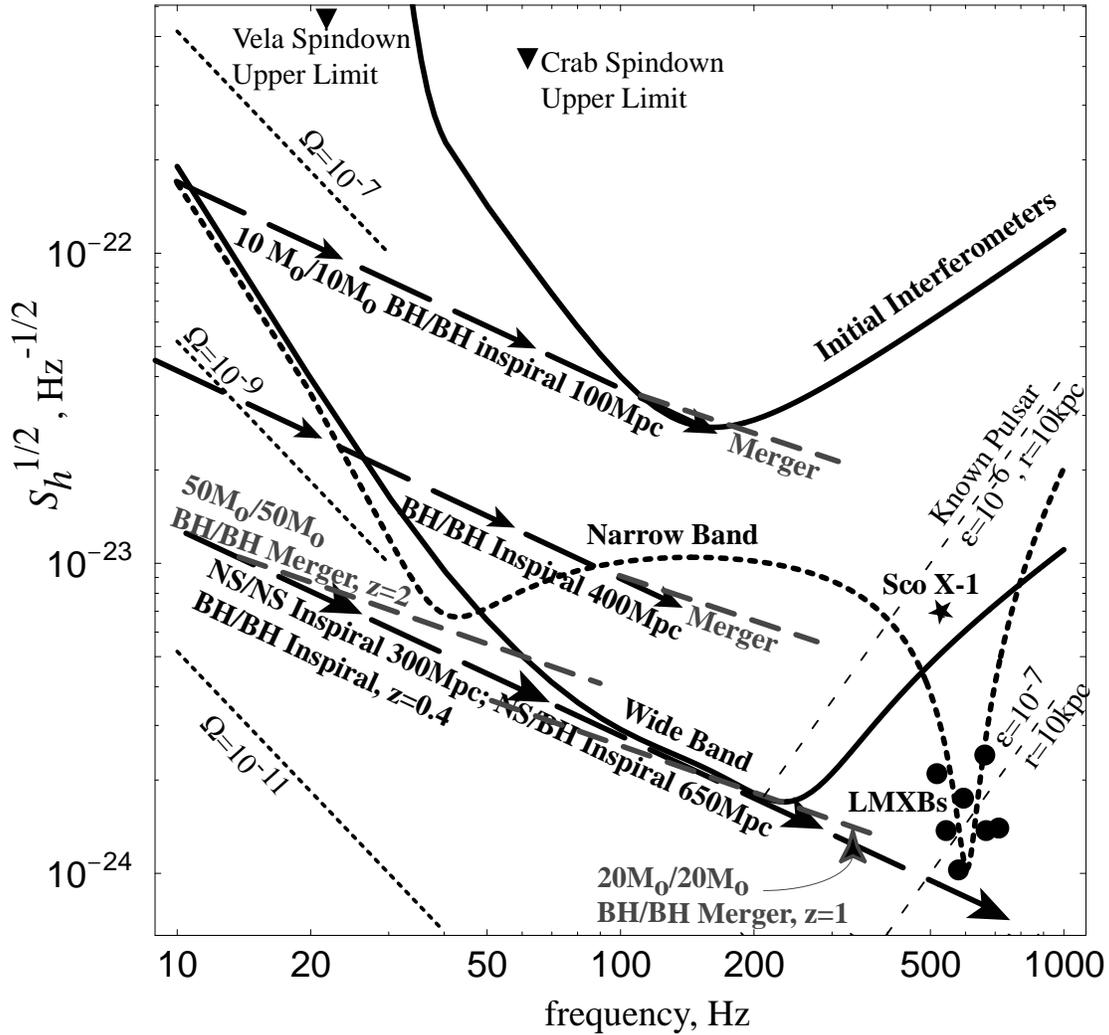}
\caption{Comparison of source strength to noise magnitude for several
astrophysical gravitational-wave sources.  The heavy black bands are
the various incarnations of LIGO interferometers: initial (top,
solid), advanced in wide band configuration (bottom, solid), advanced
in narrow band configuration (bottom, dotted).  The meaning of the
various source markers included here is described in the text.
(Figure adapted by Kip Thorne for our use, from Ref.\
{\cite{ligo_mature}}.)}
\label{fig:ligo_srcs}
\end{figure}

Figure {\ref{fig:ligo_srcs}} illustrates several important
gravitational-wave sources compared with LIGO noise curves.  The noise
curves shown are the amplitude spectrum ${\tilde h}$, with units ${\rm
Hz}^{-1/2}$, for first generatation detectors, and for advanced
detectors in both wide and narrow band configurations
{\cite{explain_sensitivity}}.  For each source, the plotted strain
${\tilde h}_s$ is defined such that the ratio is equal to the ratio of
signal strength $S$ to noise threshold $T$, rms averaged over all
source directions and orientations: ${\tilde h}_s/{\tilde h} = \langle
S^2/T^2 \rangle^{1/2}$.  The threshold is set assuming that the best
practical data analysis techniques have been applied, and that the
probability of a false alarm is one percent.  Thus, a source that
touches the noise in this figure is at the threshold for detection.
Details of how the threshold is computed for each source are discussed
in the following subsections.

\subsubsection{Compact binaries}

Coalescing compact binary star systems are currently the best
understood sources of gravitational waves.  Members of such binaries
are compact, collapsed stellar remnants --- neutron stars or black
holes.  Double neutron star binaries are observed in our galaxy;
detailed studies of these systems
{\cite{hulse_taylor,taylor_weisberg,stairsetal}} provide what is
currently our best data on gravitational-wave emission and led to
Joseph Taylor and Russell Hulse winning the Nobel Prize in 1993.  The
galactic binary neutron star systems have orbital periods of several
hours, radiating at $\mbox{several}\times10^{-4}\,{\rm Hz}$, far
beyond the range of ground-based detectors.  Gravitational waves carry
energy and angular momentum away from the system, driving the neutron
stars to inspiral towards one another.  The emitted waves sweep
upwards in frequency and amplitude as the neutron stars come closer
together.  In several hundred million years, the gravitational waves
from a galactic binary will enter the frequency band of LIGO
detectors; about 10 minutes later, their neutron stars will violently
collide and merge into a single object.

A few hundred million years is a bit long to run an experiment, so
gravitational-wave detectors must be sensitive to a large volume of
the universe in order to obtain an interesting event rate.
Extrapolating from the observed population to the universe at large
(see, e.g., Refs.\ {\cite{nps,phinney91,kalogera_lorimer}}) suggests
that in order to measure an interesting rate of events (several per
year), detectors must be sensitive to coalescences hundreds of
megaparsecs away.  Extrapolation can only predict the rate of binary
neutron star systems, since those are the only compact binaries that
have been observed to date.  Another method is population synthesis:
modeling stellar evolution to predict the rate of compact binary
formation from a population of main sequence stars (see, e.g., Refs.\
{\cite{bethebrown,fwh,belczynbulik,pzy,pzm}}).  Population synthesis
predictions span a rather wide range --- not surprising, since much of
the underlying physics is rather uncertain.  Most calculations agree
at least to within an order of magnitude with the extrapolated
predictions for binary neutron star coalescence.  This is by design:
the calculations are ``tuned'' to match the data for binary neutron
stars.  Predictions for other compact binary systems, neutron
star-black hole and black hole-black hole, vary quite a bit
{\cite{bethebrown,pzy,pzm}}.  Good data from gravitational-wave
detectors will have a large impact on our understanding of stellar
evolution and compact binary formation.

The entire coalescence process can be usefully (albeit somewhat
crudely) divided into three broad epochs {\cite{flanhughes}}: the {\it
inspiral}, in which the binary's members are widely separated and
spiral inward due to gravitational-wave backreaction; the {\it
merger}, in which the binary's orbit becomes dynamically unstable, and
the two bodies merge into a single object; and (possibly) the {\it
ringdown}, in which the merged remnant settles down to a stationary,
rotating (Kerr) black hole.  The ringdown only occurs if the binary
leaves a black hole behind after coalescence.  This broad
characterization is rather oversimplified, but useful, providing a
qualitative description of the system's dynamics and wave emission.
Further discussion of these epochs in the context of current research
on coalescing binaries is given in Sec.\ {\ref{sec:sciencereach}}.

Dividing the system's evolution into three epochs likewise divides the
gravitational-wave signal into three broad frequency bands.  This is
one reason that this crude characterization is useful: it gives some
sense of what source dynamics are accessible to the observatories.
Consider first the inspiral.  Very roughly speaking (cf.\ discussion
in Sec.\ IIIB of Ref.\ {\cite{flanhughes}}),
\begin{equation}
\mbox{inspiral waves:}\qquad f \alt 400\,{\rm Hz}
\left[10\,M_\odot\over(1 + z)M\right]\;.
\label{eq:insp_freq}
\end{equation}
(Here, $z$ is the cosmological redshift.)  Likewise, we can with a
fair degree of confidence characterize the frequency band of the
ringdown waves.  Using black hole perturbation theory
{\cite{teuk73,leaver}}, we find that the mode that is most likely to
dominate after binary coalescence has
\begin{eqnarray}
\mbox{ringdown waves:}\qquad f &\sim& {c^3\over 2\pi G(1 + z)M}
\left[1 + 0.63(1 - a/M)^{0.3}\right]\nonumber\\
&\sim& (1200 - 3200)\,{\rm Hz}\left[10\,M_\odot\over(1 + z)M\right]\;.
\label{eq:ring_freq}
\end{eqnarray}
The Kerr spin parameter $a$ varies from 0 to $M$; the span in
frequency reflects this range.  Merger waves are then all waves
emitted at frequencies between these two extremes.

Note that all frequencies scale inversely with the redshifted mass,
$(1 + z)M$.  The redshift factor should be fairly obvious, since $1 +
z$ is precisely the factor by which radiation's wavelength changes due
to cosmological evolution.  The inverse mass dependence enters because
the mass sets all timescales relevant to the radiation.  Since
frequency is an inverse timescale, all frequencies are proportional to
inverse mass.

As mentioned in the discussion of the LDAS architecture, some systems
will be analyzed using {\it matched filtering}, correlating the
experimental data with theoretical waveform models, known as templates
{\cite{cutflan,ben}}.  Matching a template to the signal boosts the
signal-to-noise ratio (SNR) by a factor that is roughly $\sqrt{N}$,
where $N$ is the number of measured gravitational-wave cycles.
Because the ringdown and early inspiral are rather well-understood, we
are confident in our templates for those coalescence epochs; the late
inspiral and merger epochs are rather more poorly understood.  This
motivates much research in relativity theory today.

The tracks illustrating compact body coalescence in Fig.\
{\ref{fig:ligo_srcs}} assume that matched filtering is applied.  The
thresholds are computed assuming that matched filters are used to
integrate the signal at each frequency over bands of width $\Delta
f\sim f$.  We assume that data from all three LIGO interferometers is
combined.  For neutron star-neutron star binaries, only the inspiral
waves are easily accessible to ground-based detectors.  As will be
mentioned in Sec.\ {\ref{subsec:dense_nuclear}}, narrow banded
interferometers and acoustic detectors will be able to provide some
information about the high frequency binary neutron star merger
signal.  The binary neutron star signal is fairly weak; for assured
detection by first detectors (amplitude SNR $\sim 5$ in all LIGO
interferometers), the binary can be no further than about 20 Mpc.  We
must go much farther out to obtain an interesting rate of neutron star
coalescences.  As a consequence, current wisdom is that neutron
star-neutron star detection by first detectors, though plausible, is
not likely.  However, detection after upgrade is quite likely --- by
Figure {\ref{fig:ligo_srcs}}, neutron star-neutron star coalescence
should be seen out to about 300 Mpc.  Non-detection of binary neutron
star coalescence by advanced interferometers would be rather
surprising.

Neutron star-neutron star waves are weak because of the system's
relatively small mass.  Increasing the mass increases the signal, but
shifts all frequency bands downward.  The shift in frequency causes
much of the inspiral to shift out of band, leaving the very late stage
of inspiral and the merger in LIGO's most sensitive band.  Figure
{\ref{fig:ligo_srcs}} shows that for the more massive binary black
hole systems, the late inspiral and merger waves are likely to be the
most relevant for detection.  The fact that these rather poorly
understood epochs of binary black hole coalescence are right in the
most sensitive band of ground-based detectors greatly motivates
theoretical work to understand this waves better.  If we succeed in
modeling the late inspiral and merger accurately, these signals should
be detectable at cosmological distances, with redshift $z \sim 0.5 -
1$.

\subsubsection{Stochastic backgrounds}
\label{subsubsec:ligo_stochastic}

Stochastic backgrounds are ``random'' gravitational waves, arising
from a large number of independent, uncorrelated sources that are not
individually resolvable.  A particularly interesting source of
backgrounds is the dynamics of the early universe --- an all sky
gravitational-wave background, similar to the cosmic microwave
background.  Backgrounds can arise from amplification of primordial
fluctuations in the universe's geometry, phase transitions as
previously unified interactions separated, or the condensation of a
brane from a higher dimensional space.  Mechanisms of this kind and
their connections to unification physics are discussed in Sec.\
{\ref{subsec:new_physics}}; here, we briefly discuss how these
backgrounds are characterized and the levels of sensitivity LIGO
should achieve.

Stochastic backgrounds are always discussed in terms of their
contribution to the universe's energy density, $\rho_{\rm gw}$.  In
particular, one is interested in the energy density as a fraction of
the density needed to close the universe, over some frequency band:
\begin{equation}
\Omega_{\rm gw}(f) = {1\over\rho_{\rm crit}}{d\rho_{\rm gw}\over
d\ln f}\;.
\label{eq:stoch_gw_om}
\end{equation}
Different cosmological sources produce different levels of
$\Omega_{\rm gw}(f)$, centered in different bands.  Amplified
primordial fluctuations surely exist, but are likely to be rather
weak: estimates suggest that the spectrum will be flat across LIGO's
band, with magnitude $\Omega_{\rm gw} \sim 10^{-15}$ {\cite{turner}}.
Waves from phase transitions can be significantly stronger, but are
typically peaked around a frequency that depends on the temperature
$T$ of the phase transition:
\begin{equation}
f_{\rm peak} \sim 100\,{\rm Hz}\left({T\over10^5\,{\rm TeV}}\right)\;.
\label{eq:freq_phasetrans}
\end{equation}
We note here that the temperatures required to enter the LISA band,
$f\sim 10^{-4} - 10^{-2}$ Hz, is $T \sim 100 - 1000\,{\rm GeV}$,
nicely corresponding to the electroweak phase transition.  Waves
arising from extradimensional dynamics should peak at a frequency
given by the scale $b$ of the extra dimensions
{\cite{hogan_prl,hogan_prd}}:
\begin{equation}
f_{\rm peak} \sim 10^{-4}\,{\rm Hz}\left({1\,{\rm mm}\over
b}\right)^{1/2}\;.
\label{eq:freq_extradim}
\end{equation}
For the waves to be in LIGO's band, the extra dimensions must be
rather small, $b \sim 10^{-15}$ meters.  LISA's band is accessible for
a scale similar to those discussed in modern brane-world work
{\cite{randallsundrum1,randallsundrum2}}.

The detectable magnitude of $\Omega_{\rm gw}(f)$ and the sensitive
frequency band depend on the particular instrument.  LIGO will measure
stochastic backgrounds by comparing data at the two sites and looking
for correlated ``noise'' {\cite{maggiore,allen_romano}}.  For
comparing to a detector's noise, one should construct the
characteristic stochastic wave strain,
\begin{equation}
h_c \propto f^{-3/2} \sqrt{\Omega_{\rm gw}(f) \Delta f}\;.
\end{equation}
(For further discussion and the proportionality constants, see
{\cite{maggiore}}.)  Stochastic backgrounds are illustrated in Fig.\
{\ref{fig:ligo_srcs}} as the downward sloping dotted lines, labeled
$\Omega = 10^{-5}$, $\Omega = 10^{-7}$, etc.  These curves assume
$\Delta f \sim f$.  Early detectors will have fairly poor sensitivity,
sensitive to a background at levels $\Omega_{\rm gw}\sim
5\times10^{-6}$ in a band from about 100 Hz to 1000 Hz.  This is
barely more sensitive than known limits from cosmic nucleosynthesis
{\cite{allen_review}}.  Later upgrades will be significantly more
sensitive, able to detect waves with $\Omega_{\rm gw}\sim 10^{-10}$,
which is good enough to place interesting limits on cosmological
backgrounds.

\subsubsection{Stellar core collapse}

Another promising source of radiation for ground-based detectors is
the core collapse of massive stars.  When massive stars die their most
dense central regions catastrophically collapse, driving the star to
explode in a supernova and the innermost matter concentrations to form
a neutron star or black hole.  The conditions for strong gravitational
wave emission --- highly energetic dense matter dynamics --- are
clearly met.  However, because stellar collapse is not very well
understood, our understanding of gravitational-wave emission from such
collapse is rather uncertain.

One of the major goals of gravitational-wave observatories is to
coordinate gravitational-wave detection with external observations.
This will be particularly important for stellar core collapse due to
our poor understanding of the likely gravitational waves emitted in
that case --- coincident detection with neutrino, gamma ray, and/or
optical observatories will greatly increase confidence in the
measurements, as the combined network of instruments provides triggers
and cross-checks for each other.  Coincident detection and measurement
will produce great science payoff aside from increased confidence and
cross-checks.  For example, a supernova in our galaxy would be easily
detected by both neutrino and gravitational-wave observatories.
Measurement of such events will paint a far more solid picture of
collapse dynamics than our current view, tracking the relative
evolution of processes that generate neutrinos with the dense matter
dynamics.  An event in our galaxy would have very large SNR, so such
studies could be done with high precision.  Such nearby events will be
very rare (the supernova rate is estimated to be several events per
galaxy per century), but would provide enormous scientific payoff.

A review of gravitational waves from stellar collapse, focusing on
their relevance to gravitational-wave observatories, has recently
appeared {\cite{fhh}}; our discussion is largely based on that paper.
The strongest core collapse gravitational waves come from
instabilities that can develop in the matter dynamics.  One of the
most-discussed instabilities is {\it bar formation}: the tendency of
rotating matter to go into a non-axisymmetric, rotating, bar-shaped
mode.  A bar mode is potentially a strong source of radiation.  Many
numerical simulations
{\cite{tohline85,durisen86,williams88,houser94,smith96,houser96,pickett96,houser98,new00}}
have shown that bar modes are promising sources, though usually with
rather simplified models.  Ref.\ {\cite{fhh}} looked at the potential
for bar mode instability in a variety of realistic stellar collapse
scenarios, and found that it still remains a promising source of waves
for ground-based observatories, though much work remains to be done to
clarify the waves' characteristics in different circumstances.  For
example, bar modes will only be unstable if the precollapse progenitor
is rotating sufficiently rapidly.  This appears likely for many
supernova progenitors.  It {\it may} also be the case in ``accretion
induced collapse'' (AIC) of some white dwarf stars, though the
conditions for this to occur and its likely rate make AIC much less
promising as a source {\cite{lindblomliu,liu,liu_private}}.

In some star collapses, the distribution of density and angular
momentum is such that the dense inner material may fragment into
pieces, forming ``chunks'' that rapidly orbit for some time before
settling into an equilibrium.  These orbiting chunks would be copious
gravitational-wave radiators.  Van Putten has recently argued that
such a matter distribution is likely to form in the progenitor to a
gamma-ray burst (cf.\ Ref.\ {\cite{mhvp}} and references therein).
Gamma-ray bursts may thus be accompanied by an extraordinarily strong
burst of gravitational radiation.  Fryer, Holz, and Hughes
{\cite{fhh}} find that a fragmentation instability may allow
ground-based detectors to observe the collapse of population III stars
--- a putative population of extremely massive ($\sim 100 -
300\,M_\odot$) stars that formed and died early in the universe's
history.  This is an example of new astrophysics that
gravitational-wave astronomy may discover.

\subsubsection{Periodic sources}

Periodic sources of gravitational waves are emitters which radiate at
constant frequency (or nearly constant frequency), much like radio
pulsars.  At first sight, this suggests they might be easy to detect
--- it should be simple to coherently follow a periodic source's phase
evolution, allowing us to build up enough power for an intrinsically
weak signal to stand above noise.  However, the signal is strongly
modulated by the earth's rotation and orbital motion, ``smearing'' the
wave across multiple frequency bands and greatly degrading the
source's strength.  Searching for periodic gravitational waves means
demodulating the motion of the detector.  This is computationally
arduous --- the modulation is different for every sky position.
Unless one knows in advance the position of the source, one needs to
search over a huge number of sky position ``error boxes'', perhaps as
many as $10^{14}$.  One rapidly becomes computationally limited.  For
further discussion, see {\cite{periodic}}; for ideas about doing
hierarchical searches, requiring less computational horsepower, see
{\cite{per_hierarchical}}.

The prototypical source of continuous gravitational waves is a
rotating neutron star.  If the neutron star is non-axisymmetric (for
example, it has a crust that is somewhat oblate and misaligned with
the star's spin axis), it will radiate gravitational waves with
characteristic amplitude
\begin{equation}
h_c \sim {G\over c^4}{I f^2 \epsilon \over r}\;,
\end{equation}
where $I$ is the star's moment of inertia, $\epsilon$ characterizes
the degree of distortion (see {\cite{periodic}} and references therein
for further discussion), $f$ is the wave frequency, and $r$ is the
distance to the source.  Realistic choices for these parameters
suggest that $h_c \sim 10^{-24}$ or smaller.  To measure these waves,
we need to coherently track the signal for a large number of wave
cycles.  Coherently tracking $N$ cycles boosts the signal strength by
a factor $\sim\sqrt{N}$.

The characteristic amplitude of several possible pulsars are
illustrated in Fig.\ {\ref{fig:ligo_srcs}}.  The points labeled
``Crab'' and ``Vela'' illustrate the maximum possible
gravitational-wave strength that the well-known Crab and Vela pulsars
could emit.  This pulsars are known to ``spin down''; that is, their
rotation periods are observed to decrease over time.  Figure
{\ref{fig:ligo_srcs}} assumes that {\it all} of this spindown is due
to gravitational-wave backreaction, an assumption that is known to be
wrong.  The waves from these pulsars will be much weaker; however,
since their sky position and frequencies are known in advance, it will
not be too difficult to search for any waves they might emit.  The
upward sloping lines labeled ``Known pulsar, $\epsilon = 10^{-6},
10^{-7}$, $r = 10\,{\rm Kpc}$'' illustrate the waves produced by a
source whose position is known but frequency is unknown, if the
distance and deformation are as indicated.  In all of these cases, it
is assumed that the signal can be coherently tracked for several
months, building up power.

A nearly periodic source of gravitational waves from rotating neutron
stars is the r-mode instability, a fluid instability that may be
active in hot, fluid neutron stars
{\cite{andersson,friedman_morsink,lom,aks}}.  This mode is not
perfectly periodic, since radiative backreaction rapidly reduces the
star's spin and gravitational-wave frequency; but it is nearly so, and
search techniques for r-mode active stars are likely to be very
similar to searches for periodic sources {\cite{olcsva}}.

Finally, we note that neutron stars accreting matter from a close
companion may be periodic sources of gravitational radiation
{\cite{lars,ubc}}.  This would explain why most neutron stars in such
low mass x-ray binary (LMXB) systems appear to have a maximum spin
frequency: rather than spinning up indefinitely by the matter
accreting onto them, they eventually are braked by gravitational-wave
backreaction.  Such sources are very promising targets for observation
since their sky positions are known very accurately from x-ray
observation.  Since they are in binaries, though, their waveforms will
be even more strongly modulated than ``normal'' radiating pulsars.
The points on Fig.\ {\ref{fig:ligo_srcs}} labeled ``Sco X-1'' and
``LMXBs'' show waves that are possible from known LMXB systems.  Many
of these systems will only be detectable in the narrow-banded
configuration, demonstrating the science impact possible with tunable
interferometer configurations.

\section{Space-based detectors}
\label{sec:spacebased}

Many promising sources of gravitational waves radiate at low
frequencies, from a few microHz to about a Hz, where Earth-based
detectors have very poor sensitivity due to geophysical noise.  Such
sources include massive black hole systems, compact binaries, very
close normal binaries, and cosmological backgrounds.  To get away from
the geophysical background, the detector must be located in space.  In
this section, we describe the space-based detector LISA which will
target these low-frequency sources.  We first describe the LISA
mission, and then discuss certain sources that are of particular
interest for LISA science.  More extensive information on the mission
and additional references are given in Refs.\
{\cite{prephaseA,intlisa_1,intlisa_2,bender}}.

\subsection{Description of the LISA Mission}
\label{subsec:lisa_mission}

The Laser Interferometer Space Antenna (LISA) is designed to detect
and study in detail gravitational-wave signals at frequencies from
roughly 10 microHz to 1 Hz.  Once the conclusion to build a
low-frequency, space-based detector has been reached, it becomes
attractive to make the size of the antenna quite large.  Using laser
heterodyne measurements between widely separated spacecraft, even with
1 W power levels, antenna sizes of millions of kilometers are
feasible.  The measurements are made between freely floating proof
masses, inside the spacecraft, that are very carefully shielded from
both internal and external disturbances.  With this approach, very
desirable sensitivity levels can be achieved throughout the planned
frequency range.

The LISA mission is planned as a joint mission of the European Space
Agency and NASA, for launch in about 2010.  The antenna will be based
on laser measurements between three spacecraft located at the corners
of an equilateral triangle that is 5 million km on a side.  A one year
orbit around the Sun for each spacecraft has been found such that: (a)
the spacecraft separation stays constant to better than $1\%$ over
more than 10 years; (b) the center of the antenna is on a circular
orbit in the ecliptic plane and about 50 million km ($20^\circ$)
behind the Earth; (c) the plane of the antenna is tipped at $60^\circ$
to the ecliptic, and that plane rotates around the pole of the
ecliptic with a one year period; and (d) the antenna rotates within
that plane with a one year period.  The orbits each have an
eccentricity of about 0.01, and an inclination that is the square root
of 3 times the eccentricity.

\begin{figure}[t]
\includegraphics[width = 15cm]{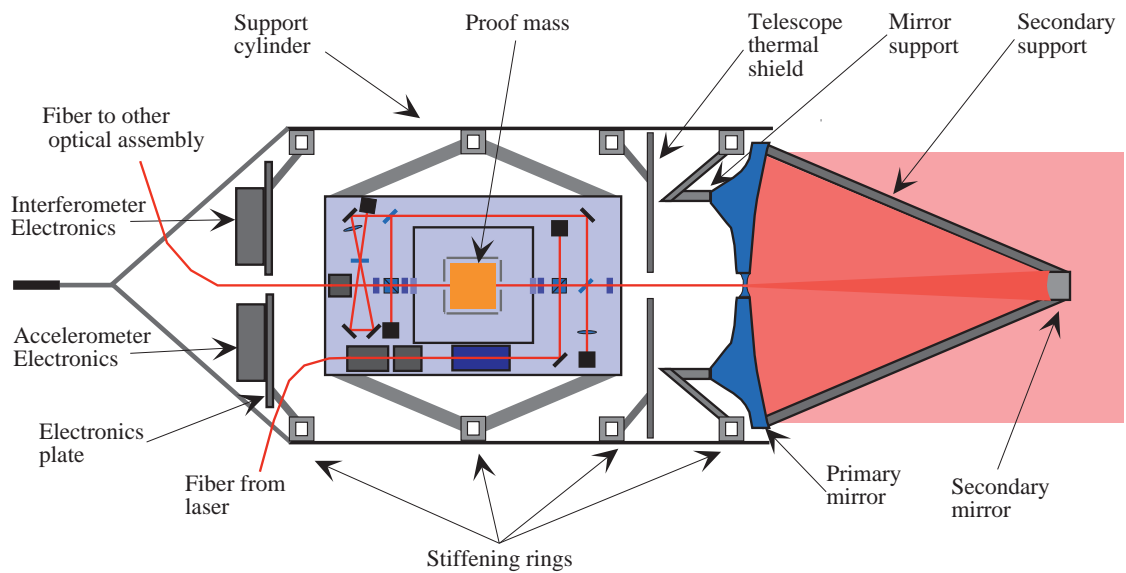}
\caption{Schematic of the LISA optical assembly.}
\label{fig:opt_assembly}
\end{figure}

The main components of the payload for each spacecraft are located
within a Y-shaped thermal shield made of carbon fiber reinforced
plastic, which has a low coefficient of thermal expansion.  There are
two cylindrical optical assemblies about 1 m long and 0.35 m in
diameter that point out along two of the arms of the Y and toward the
distant spacecraft.  A schematic drawing of one optical assembly is
shown in Fig.\ {\ref{fig:opt_assembly}}.  Only a brief description of
its operation will be given here.

The rectangular structure located near the center of the optical
assembly is the optical bench.  It is supported by a Wheatstone bridge
mounting from the optical assembly support cylinder so that
temperature gradients along that cylinder don't get transmitted
conductively to the optical bench.  The gravitational sensor is
mounted at the center of the optical bench, and contains the freely
floating proof mass and the housing around it.  Capacitive bridge
measurements between a number of pairs of capacitive plates on the
inside of the housing and the proof mass determine even small
translations and rotations of the proof mass with respect to the
housing.  The laser measurements of changes in the distances between
the proof masses in different spacecraft are made to the front face of
each proof mass through a small window in the outer enclosure of the
gravitational sensor.

Light from the 1.06 micron Nd:YAG laser (not shown) is brought onto
the optical bench through a fiber, and almost all of it is sent by a
polarizing beamsplitter to the 0.3 m diameter transmit/receive
telescope.  The transmitted beam has a diameter of about 20 km when it
reaches the distant spacecraft, and a small portion of it is collected
by the telescope there.  This light goes through the beamsplitter, is
reflected from the face of the proof mass, and is then deflected by
the beamsplitter to go to a photodiode detector.  Some light from the
on-board laser also goes to the photodiode, and the beat between the
received beam and this local laser beam is one of the six main signals
generated by the antenna.  Each beam propagating in one of the two
directions along one of the three sides of the antenna is used in
generating one of the six main signals.

Because of the choice of spacecraft orbits, the lengths of the sides
of the triangle stay constant in length to better than $1\%$ over 10
years or longer, even including planetary perturbations.  However, the
resulting relative velocities are still on the order of 10 m/s, and
cause Doppler shifts of up to roughly 10 MHz at some times.  These
Doppler shifts vary typically with periods of several months or longer
and in an extremely smooth fashion, so that they can be fitted out
from the observations.  The gravitational waves will cause phase
variations in the LISA signals that can be analyzed after the huge but
much lower frequency Doppler shifts have been fitted out.

To understand how the LISA antenna works, it is useful to consider
only two sides of the triangle, with the spacecraft common to both
sides being the master spacecraft.  The lasers at the two distant
spacecraft are assumed to be phase locked to the received laser beams,
so that the return beams have the same phase as if the beams were just
reflected by mirrors.  Thus the system acts like a Michelson
interferometer, but with two signals generated back at the master
spacecraft that give the variations in the length of each of the
interferometer arms separately.  To the extent that the proof masses
at the ends of the arms really are undisturbed by spurious
accelerations, the apparent variations in the sum of the lengths of
the two arms will almost all be due to phase noise in the laser.  This
information then can be used to correct the difference in length of
the two arms in software for the laser phase noise, which otherwise
would swamp the gravitational wave signals.

An Industrial Phase A Study of the LISA mission was carried out for
the European Space Agency and presented to their scientific community
in September 2000.  The spacecraft, payload and mission designs from
this study, which started from the results of earlier studies in both
Europe and the US, essentially form the current baseline plan for the
mission.  Under this plan, each spacecraft is 2.7 m across and 0.56 m
high, with the sides slanted in at $30^\circ$ so that sunlight only
hits the top surface, where the solar cells to provide power are
mounted.  The optical assemblies point out through the slanted sides,
and the direction to the Sun makes a constant $30^\circ$ angle with
respect to the top of the spacecraft.  This keeps the temperature and
temperature distribution in the spacecraft extremely constant.  This
is a major reason why the time variations in spurious forces acting on
the carefully shielded proof masses deep in the spacecraft can be kept
extremely low.

Each spacecraft initially has a propulsion module attached to it, and
the three combined units are stacked up and launched by a single
launch vehicle into roughly 13 month period elliptical orbits around
the Sun.  After leaving the Earth, the units separate and their
propulsion units carry them to their desired orbits over a period of
roughly a year.  The orbits are checked for a few weeks by tracking
with NASA's Deep Space Network, and minor orbit corrections are made
as needed.  The propulsion modules then separate gently from the
spacecraft and drift away.  After separation, the spacecraft are
oriented properly with their optical assemblies pointing at the other
spacecraft forming the antenna.  This is done using continuously
operating and electrically controllable micronewton thrusters mounted
on the spacecraft.

Next, the proof masses are released from the clamping mechanisms that
have held them during launch and electrical forces are applied to move
them to the centers of their housings.  Then, the control laws for the
micronewton thrusters are changed so that they also take on the job of
making the spacecraft follow the average position of the two proof
masses.  The motion of the spacecraft is then determined by the forces
acting on the proof masses, and the effect of non-gravitational forces
on the spacecraft can be essentially eliminated.  This is called a
drag-free system, since such systems were first used to greatly reduce
the drag on Earth satellites due to the residual atmosphere.  For
LISA, since there are two proof masses, weak electrical forces
perpendicular to the laser beam directions are still applied to the
proof masses to keep them centered in their housings.

The capacitive position measurements are sensitive enough and the
noise in the thrusters low enough that the relative motion of the
spacecraft with respect to the proof masses can be kept very low.
This is another reason why the variations in spurious forces on the
proof masses can be kept very low.  The total thrust needed is mainly
the roughly 20 micronewtons required to buck out the solar radiation
pressure, and varies only slightly.

Finally, a beam acquisition procedure is started to successively turn
on each laser and adjust its pointing direction accurately.  When all
six main signals are obtained, the scientific part of the mission can
begin.  The nominal scientific mission lifetime may be as short as two
or three years, but it is hoped that the observations can continue for
about a decade.  The thrust required from the micronewton thrusters is
low enough that the reaction mass needed even for a decade of
operation is small.

\subsection{LISA Sensitivity and Galactic Binaries}
\label{subsec:gal_binaries}

\begin{figure}[t]
\includegraphics[width = 17cm]{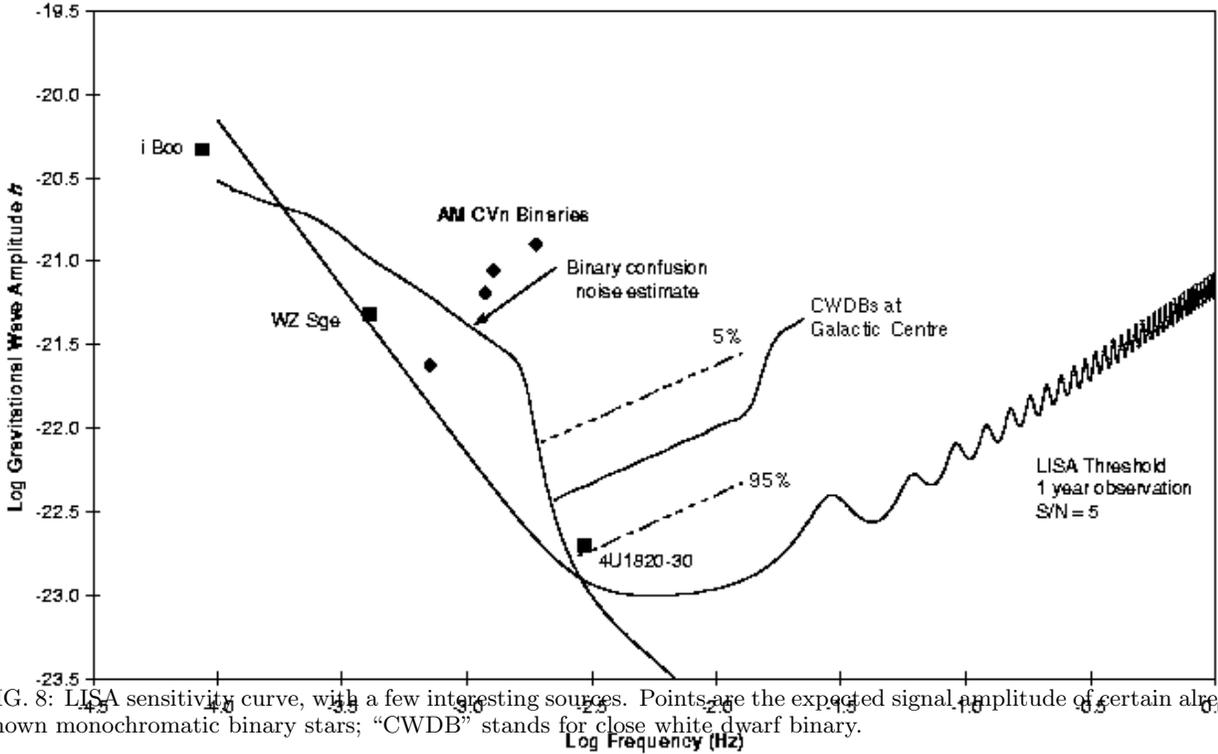}
\vskip -2cm
\caption{LISA sensitivity curve, with a few interesting sources.
Points are the expected signal amplitude of certain already known
monochromatic binary stars; ``CWDB'' stands for close white dwarf
binary.}
\label{fig:lisa_sens}
\end{figure}

The instrumental sensitivity of the LISA antenna is determined by two
factors.  One is the noise in measuring changes in the distances
between the different proof masses.  The second is the level of
spurious accelerations of the proof masses.  From noise budgets for
these two different types of noise, a threshold sensitivity curve for
LISA can be determined.  Such a curve is shown in Fig.\
{\ref{fig:lisa_sens}}, along with some information about expected
signals from binary stars in our galaxy that will be discussed later.
The sensitivity curve shown is for 1 year of observations and an
amplitude SNR of 5.

Nearly all of the resolvable galactic binary sources expected for LISA
will be thousands to millions of years from coalescence, and thus
their frequencies will change little over a few years.  Each can be
represented by a single point giving the rms gravitational-wave
amplitude and frequency (twice the orbital frequency).  If that point
lies above the sensitivity curve, then there is enough signal to
overcome the instrumental noise with 1 year of observations and a SNR
of 5.  This SNR is needed to be confident of detection for a source
whose sky position and frequency are not known.

There is of course an error allocation budget for each of the two
types of noise.  In such a budget, allowable levels of error are
assigned to each of the expected or possible sources of error in the
measurements.  For measuring changes in the distances between proof
masses, the total error budget level for the round-trip arm length
difference of two antenna arms is 40 picometers per root Hz (pm/rtHz),
independent of frequency (white noise) down to below 1 millihertz (1
mHz).  This is called the spectral amplitude of the noise, and is
defined as the square root of the power spectral density.  The noise
power in a narrow bandwidth is proportional to the bandwidth, with
units of [(pm)$^2$]/Hz, so the spectral amplitude has units of
pm/rtHz.

Most of the expected distance measurement noise comes from shot noise
in the detected laser photons.  The rate of photon detections for the
received signal is a few times $10^8$ per second, so measurements can
be made to roughly $1/(2\pi\sqrt{10^8})\sim 10^{-5}$ of a laser
wavelength in 1 second, or about $10^{-21}$ in 1 second for the
fractional change in the round-trip distance.  There is some
additional contribution to the error due to jitter in the pointing of
the laser beam toward the distant target and other noise sources, but
they are relatively small.

The LISA sensitivity curve at frequencies substantially above 3 mHz is
determined by the distance measurement noise.  At lower frequencies,
the instrumental sensitivity is limited mainly by spurious
accelerations of the proof masses.  There are roughly half a dozen
sources of such acceleration noise that dominate the error budget, and
they are assigned equal error levels.  With allowances for other
smaller acceleration noise sources, the rms total is $3\times10^{-15}$
meters sec$^{-2}$/rtHz down to 0.1 mHz, and somewhat higher at still
lower frequencies.  A space validation flight for the subsystem
consisting of the proof mass, its housing, and the associated
electronics currently is planned in order to verify that they perform
according to their design specifications.

With the sensitivity curve shown in Fig.\ {\ref{fig:lisa_sens}}, LISA
will be able to observe signals from many binaries in our galaxy.  In
order for the two members of a binary to be close enough together to
have gravitational wave frequencies above about 0.1 mHz, both must be
compact objects --- white dwarf stars, neutron stars, or black holes.
The number of such binaries in our galaxy is roughly $10^8$; most will
be close white dwarf binaries.  Below 1 or 2 mHz there will be more
than one in each 1 cycle/year frequency resolution bin, so that
individual ones cannot be resolved.  The curve labeled ``confusion
noise estimate'' in Fig. {\ref{fig:lisa_sens}} represents the expected
noise level due to the random superposition of such signals.

At higher frequencies, the number per bin drops rapidly with
increasing frequency, as they lose energy more and more rapidly by
gravitational-wave emission.  About 3000 will be resolvable above
about 3 mHz.  Roughly $90\%$ of them will have amplitudes between the
curves labeled $5\%$ and $95\%$ in the figure.  The direction and
frequency of only a few of them will be known ahead of time, and some
of these are shown.

\subsection{Massive Black Hole Binaries}
\label{subsec:massive_bbh}

There are at least three types of events involving massive black holes
(MBHs) that appear likely to produce signals observable by LISA.
However, the event rates can only be estimated very crudely, and this
situation seems unlikely to change much in the near future.  The
primary objective of the LISA mission is to detect such signals and
study them in detail.  But none of these signals are guaranteed to be
present, so the present design of the LISA antenna has been chosen to
make as many of the types of MBH signals observable as possible.

The first type of signal of strong interest is associated with the
origin of massive black holes.  An important question is how the seed
MBHs that later grew to be the massive and supermassive black holes
observed today were formed.  We label black holes with masses $M\sim
1.5 - 100\, M_\odot$ ``stellar mass black holes'' (potentially formed
by the evolution of massive stars); holes with mass up to about
$3\times10^7\,M_\odot$ are labeled ``massive black holes'' (MBHs); and
still larger ones ``supermassive black holes''.

There are several theories for how seed MBHs form.  In one, stellar
mass black holes sink to the center in a dense galactic core by
dynamical friction, and collisions lead to the formation of higher
mass objects.  The largest black hole grows faster than all others,
swallowing up holes comparable size.  This becomes the seed for growth
of an MBH of perhaps $10^5\,M_\odot$ or larger.  When it reaches a
mass of roughly $10^3\,M_\odot$, it would be able to continue growing
fairly rapidly by absorption of gas in the galactic nucleus and by
tidal disruption of stars {\cite{quinshap}}.

An alternate scenario postulates a dense cloud of gas and dust,
evolving to the point where it becomes optically thick.  Radiation
pressure plus magnetic fields then prevents further fragmentation of
the cloud to form stars {\cite{haehnelt_rees}}.  At that point, if
energy and angular momentum can be dissipated fairly rapidly, several
things might happen.  One possibility is that a supermassive ``star''
with mass $M\sim 10^5 - 10^6\,M_\odot$ forms.  This would be subject
to a pulsational instability {\cite{mtw}} that drives it to collapse
into a MBH.  Another possibility is that the cloud may become dense
enough to reach the point of relativistic instability and collapse
directly to a MBH without going through the supermassive star stage.

\begin{figure}[t]
\includegraphics[width = 16cm]{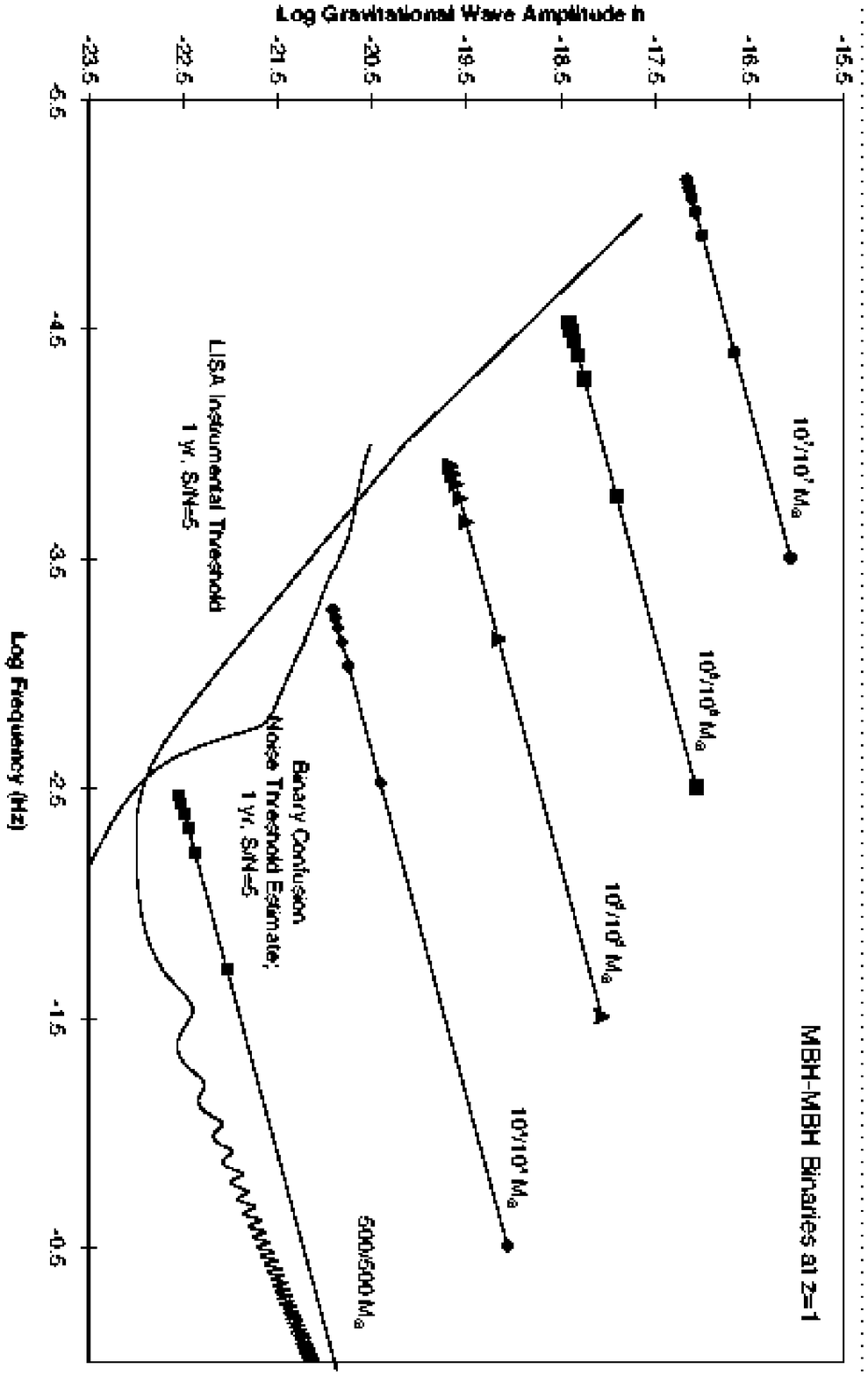}
\caption{LISA sensitivity curve compared to the wave amplitude for
binary black hole mergers.}
\label{fig:lisa_bbh}
\end{figure}

From the standpoint of LISA observations, the major issue of the
collisional growth scenario is whether some of the individual black
holes become large enough before merging with the largest hole that
their coalescence is detectable.  An example mass pair that would just
be detectable during the last year before coalescence at $z = 1$ $m_1
= m_2 = 300\, M_\odot$; another is $m_1 = 50\,M_\odot$, $m_2 = 3000\,
M_\odot$.  The lowest straight curve in Fig. {\ref{fig:lisa_bbh}}
shows the signal strength and frequency evolution during the year
before coalescence of a binary with $m_1 = m_2 = 500\,M_\odot$ at $z =
1$.  The ticks along this line represent times 0.0, 0.2, 0.4, 0.6,
0.8, and 0.99 year from the beginning of the last year.  Such a signal
would be observable even from a coalescence at $z = 5$.  The other
curves in the figure will be discussed later.

A second important signal comes from black holes of roughly
$10\,M_\odot$, neutron stars, or white dwarfs in orbit around MBHs in
galactic nuclei.  Such highly unequal mass systems will result from
compact objects in the cusp around the MBH being scattered in close
enough to start losing energy by gravitational radiation, becoming
progressively more tightly bound.  The orbits remain elliptical
throughout in this case, and during the last year the orbital speed
near periapsis can reach almost half the speed of light.  The period
for periapsis precession is similar to that for radial motion; rapid
frame-dragging may also be present.  Thus the orbits are extremely
non-Newtonian, as discussed later.  From the astrophysics point of
view, the number and type of systems of this kind observed will give
valuable information about a combination of the distribution of MBH
masses in galactic nuclei and the space density of compact objects in
the cusps around MBHs.

The third and strongest type of MBH signal expected for LISA is from
the coalescence of binaries consisting of two MBHs, where both are
substantially more massive than $10^3 M_\odot$.  Such events can
follow the mergers of galaxies or pre-galactic structures, provided
that two conditions are fulfilled.  First, the pre-merger structures
must already contain MBHs; second, the MBHs must form a close binary
in substantially less than the age of the universe.  Present estimates
are that these conditions have been fulfilled frequently enough to
give one or more MBH coalescences per year, but even this is
uncertain.  It seems likely that coalescences involving MBHs of
roughly $10^5 - 10^6 M_\odot$ will be the most common; mergers of
structures with $10^7\,M_\odot$ or more massive MBHs may well be less
frequent, and less massive MBHs may have difficulty forming close
binaries.

The signal strength during the last year for equal-mass MBH
coalescences at $z = 1$ are shown in Fig. {\ref{fig:lisa_bbh}}.  Since
the SNR ratios are high, such events would be easily observable
anywhere in the universe.  The waves from these events will be
measured sufficiently well that detailed parameter fits will be
possible {\cite{untangling}}, making possible much astrophysical
analysis.  However, since the event rate is definitely not guaranteed,
efforts will be made to keep the acceleration noise level at
frequencies below 0.1 mHz as low as possible.  This will increase the
chances of seeing coalescences or $10^5$ or $10^6\,M_\odot$ MBHs a
number of years before coalescence, and thus improve the statistical
information on them.

\subsection{LISA and stochastic backgrounds}
\label{subsec:lisa_stochastic}

By their very nature, stochastic gravitational waves are difficult to
distinguish from noise.  As described in Sec.\
{\ref{subsubsec:ligo_stochastic}}, ground-based detectors can find
them by coordinated measurements, essentially comparing the outputs of
multiple detectors to find sources of correlated noise.  Provided
there are no common noise sources (which, due to their wide
separation, should be the case), this should work well.

LISA is a single space-based antenna and so cannot use this technique,
but has other tools that can be employed instead.  In particular, it
is possible to combine the signals from LISA's spacecraft in such a
way that the detector is {\it insensitive} to gravitational waves
{\cite{armstrongetal}}.  Observables of this kind are called
``Sagnac'' observables, since they correspond to observables
constructed from phase information that propagates around the LISA
antenna, as in a Sagnac interferometer.  Since the Sagnac observable
is insensitive to gravitational waves, it measures noise alone.  In
this way, one can detangle the noise-like signature of a stochastic
background from the actual interferometer noise {\cite{hoganbender}}.
LISA's stochastic background sensitivity should then be limited by the
Galactic white dwarf background (a somewhat prosaic stochastic
background).  This will allow it to set strong limits on cosmological
backgrounds in its band, corresponding to an energy density about a
million times smaller than the cosmic microwave background.

LISA's band is fortuitously placed to measure potentially interesting
cosmological sources, as has been mentioned already in Sec.\
{\ref{subsubsec:ligo_stochastic}}.  An electroweak phase transition,
occuring when the temperature of the universe was $T \sim 100 - 1000$
GeV, generates waves peaked at a frequency $f \sim 10^{-4} - 10^{-3}$
Hz.  Such waves could be directly detected by LISA
{\cite{apredaetal}}.  Likewise, waves from extradimensional dynamics
are likely to be in LISA's band for extra dimensions on the scale of
millimeters to microns {\cite{hogan_prl,hogan_prd}}.  As discussed in
the next section, LISA is well situated to at least put stringent
limits on many early universe scenarios, and perhaps directly detect
stochastic waves.

\section{New tests of physics and tests of new physics}
\label{sec:sciencereach}

As should be clear from the discussion of sources given in Secs.\
{\ref{sec:groundbased}} and {\ref{sec:spacebased}}, the routine
detection of gravitational waves will certainly transform the field of
astronomy.  The rarest and most obscure cosmic events, such as black
hole mergers, will suddenly achieve a phenomenological prominence
appropriate to their huge power output.  Populations barely detected
at all today, such as white dwarf binaries, will produce a din that
overwhelms even the LISA instrumental noise sources.  The current
orders-of-magnitude uncertainties in rates and intensities of all
kinds of gravitational-wave sources clearly indicates our ignorance of
the dominant mass-energy flows.  We will learn about details of many
things that astrophysicists regard as fundamental: the dynamics of
massive black hole formation in the centers of galaxies, the evolution
of binary star systems, perhaps the equation of state of nuclear
matter and the mechanism of neutrino-driven supernova explosions.

Readers of these proceedings will be more interested in new physics
than new astrophysics.  These observatories should not be regarded as
``merely'' tools of astronomers; they will probe phenomena that will
either constrain or reveal something new about the laws of nature.  We
discuss these possibilities in the subsections below.  We first
describe how gravitational-wave observatories will probe the nature of
extremely strong gravity.  Section {\ref{subsec:sf_dynamic}} discusses
the gravitational-wave probes of highly dynamic gravity that will be
studied in the merger of binary black hole systems; Sec.\
{\ref{subsec:probe_bh}} focuses on measurements that will map in
detail the structure of black hole spacetimes.  It is worth noting at
this point that it is still not conclusively known whether the massive
objects usually described as ``black holes'' in fact are black holes
as described by Einstein gravity (though the evidence from sources
such as x-ray telescopes is getting better and more conclusive all the
time).  Section {\ref{subsec:dense_nuclear}} discusses how tracking
gravitational waves from systems with neutron stars may be used to
study the properties of dense nuclear matter, probing its equation of
state.  Finally, in Sec.\ {\ref{subsec:new_physics}} we examine what
kinds of ``new physics'' may be probed, focusing in particular on how
gravitational waves may be used to test new ideas for unification
physics beyond the standard model.

\subsection{Strong field, dynamic gravity}
\label{subsec:sf_dynamic}

For the general relativity theorist, some of the greatest promise of
gravitational-wave astronomy comes from the stringent tests of
classical gravity that it will provide.  The most spectacular such
tests are likely to come from observing the mergers of black holes.
Comparing the waveforms measured by ground- and space-based
observatories to those computed from theory will test the strongest
and most dynamically varying gravitational fields since the big bang.

Such comparisons obviously require us to have theoretical waveforms in
hand.  For the most interesting, extremely dynamical, strong-field
phase of binary black hole coalescences, this is unfortunately where
theory is currently stuck.  From the standpoint of mathematical
physics, this problem may appear deceptively simple: black holes
contain no matter, so one ``merely'' needs to find an appropriate
solution to the vacuum Einstein equations, $G_{ab} = 0$ (see Ref.\
{\cite{mtw}} or {\cite{wald}} for a definition of the Einstein tensor
$G_{ab}$).  Such an appropriate solution is one which: (a) consists of
two widely separated black holes in the asymptotic past; (b) consists
of a single Kerr black hole in the asymptotic future; (c) allows only
outgoing radiation to radiate ``out to infinity''; and (d) allows only
ingoing radiation to propagate down event horizons.

Anyone with experience solving nonlinear partial differential
equations subject to bizarre boundary conditions will appreciate that
this is a tall order.  At the moment, only the asymptotic past and the
asymptotic future solutions are very well understood.  When the black
holes are widely separated, one can study the binary using the
post-Newtonian approximation to general relativity.  This
approximation treats the dynamics of the system as Newtonian at lowest
order, and finds corrections to the dynamics as a power series in
$x\sim(GM/rc^2)^{1/2}$ (where $r$ is orbital radius).  This
dimensionless parameter roughly describes the gravitational field
strength.  For more detailed discussion (particularly of the many
subtleties which this discussion has ignored), see {\cite{biww}}.
Tests of general relativity in this early ``inspiral'' epoch are
essentially generalizations of current tests that use binary pulsar
systems {\cite{stta,stairsetal,damour_taylor}}.  The binary pulsars
are relatively weak-field, however, with hundreds of millions of years
remaining until the neutron stars merge.  More stringent limits on
general relativity's validity are likely to be set examining the final
few minutes of inspiral, as the members of the binary spiral through
extremely strong gravitational fields.

In the asymptotic future of the system, the holes will have merged and
formed an isolated spinning black hole.  The final waves to come from
the system will be the ``ringdown'' waves emitted as the merged
remnant settles down to the Kerr rotating black hole solution.  The
ringdown waves are those emitted by a black hole that is distorted
from its quiescent Kerr form, and can be modelled with black hole
perturbation theory {\cite{teuk73,leaver}}.  General relativity
predicts that the frequency and damping times of black hole modes
depend only on the mass and spin of the hole.  The longest lived mode
correspond to a mode with spherical-harmonic-like indices $l = m = 2$
--- a bar-like mode, rotating in the same sense as the hole's spin.
The frequency and damping times of those modes are well-understood
{\cite{echeverria}}, and so matching observation and theory should be
relatively simple.

As the inspiraling black holes come close together, their
gravitational interaction becomes progressively stronger.  Eventually,
the number of terms that must be kept in a post-Newtonian expansion
becomes large enough to be impractical.  This is the vanguard of
current research into binary systems in general relativity:
understanding how to model the dynamics and gravitational-wave
emission of extremely close binary black hole systems, particularly
the ``merger'' epoch in which the two black holes fuse into a single
object.  As discussed in Sec.\ {\ref{sec:groundbased}}, it is likely
that for some systems the most poorly understood, strong-field late
inspiral and merger regime will produce waves that are ideally suited
for detection.  This further motivates theoretical work to understand
these waves in detail.

Most efforts to model strong-field binary black hole dynamics have
focused on {\it numerical relativity} {\cite{lehner}}: large-scale
computational attempts to solve Einstein's equations, subject to the
constraints and conditions discussed above.  This has proven to be an
extraordinarily difficult problem.  Solving the 10 coupled, nonlinear
partial differential equations of the tensor equation $G_{ab} = 0$ is
difficult enough for starters.  Most efforts begin by splitting the 4
dimensions of spacetime into 3 space plus 1 time dimension.  (Not all
efforts focus on this 3 + 1 split.  There is a substantial body of
work that splits spacetime into 2 spatial directions and 2 null
directions, along which radiation propagates; see Ref.\
{\cite{twoplustwo}} and Sec.\ 3.2 of {\cite{lehner}} for detailed
discussion.)  Einstein's equations then divide into constraints which
the data must satisfy on each time ``slice'', and evolution equations
which step the data forward from slice to slice.  There is a great
deal of freedom in how one can perform the space-time split, and it is
not obvious how it can be done in a manner that is maximally useful
for both modeling a binary's dynamics and extracting the gravitational
wave information.  In addition, there are many different ways of
actually writing the Einstein equations down in preparation for a
numerical computation.  It is now clear that some forms of Einstein's
equations are particularly bad, being susceptible to the growth of
unphysical instabilities that poison a computation.  Much of the
effort in numerical relativity is now focused on finding ``good''
equations (see, for example, Ref.\ {\cite{kst}} and references
therein).  Progress in the field has been rapid recently, with large
advances in the ability to evolve systems with two black holes and
large amounts of radiation without making too many simplifying
assumptions; see Refs.\ {\cite{lazarus,lazarus_prl}} for further
discussion.

Relatively recent work by Thibault Damour and colleagues has developed
an alternative analytical approach to understanding the strong-field
dynamics of binary black hole systems; see
{\cite{buonannodamour,damour}} and references therein.  This work is
based in part on ``resummation methods'' for improving the analytic
post-Newtonian description of gravitational-wave emission, coupled
with a novel recasting of a binary's two-body dynamics in terms of the
motion of a single body in an ``effective one-body metric'', usefully
thought of as a deformed black hole.  The resummation methods are,
essentially, Pad\'e approximants, rewriting the (poorly convergent)
post-Newtonian description of the body's dynamics as a function with
improved convergence properties.  The effective one-body remapping
technique is inspired by tools that have been developed to describe
two-body problems in quantum electrodynamics; see Ref.\
{\cite{buonannodamour}} for details.  Both the new analytic techniques
and numerical relativity show great promise and will hopefully work
together to elucidate strong-field binary dynamics and their
gravitational-wave generation.

\subsection{Probing black hole spacetimes}
\label{subsec:probe_bh}

The great difficulties and uncertainties in theoretical modeling apply
to {\it comparable mass} binary black hole systems.  These systems are
also among the most interesting from the standpoint of testing
relativity, since there may be high event rates for both ground- and
space-based detectors, and because the gravitational dynamics are the
most extreme.  A special case that can be modeled with less difficulty
is that of {\it extreme mass ratio} binary black holes: one black hole
far more massive than the other.  As discussed in Sec.\
{\ref{sec:spacebased}}, these are sources that are targets for LISA.
The spacetime of the binary is describable in this case as that of a
single rotating black hole plus a perturbation arising from the small
black hole.  The system's dynamics are well-described as an
adiabatically evolving orbit of the small body.  A substantial body of
literature on such systems has developed over the past decade
(examples of which are Refs.\
{\cite{poisson,ckp,msstt,hughes_circI,hughes_circII,kenn_glam}}); all
work on this subject is based, at heart, on Teukolsky's 1973
description of radiative perturbations to rotating black holes
{\cite{teuk73}}.

These extreme mass ratio inspirals promise to be wonderful tools for
precision tests of gravitation.  Rather than probing the violent
strong-field dynamical properties found when comparable mass holes
merge, extreme mass ratio inspirals will make it possible to {\it
map}, with high accuracy, the structure of the large black hole's
spacetime.  Einstein's theory of gravitation predicts that black holes
have event horizons and have a structure that depends only on the mass
and spin of the black hole.  Extreme mass ratio inspirals promise an
extremely clean test of this structure, demonstrating once and for all
that the massive objects seen in many galaxies are ``Einstein black
holes'', or in fact are something even more bizarre.

For an astrophysically interesting range of black hole masses (about
$10^5 - 10^7\,M_\odot$ --- black hole masses commonly encountered in
the centers of galaxies {\cite{bhdemo}}) the gravitational waves
generated during inspiral are at frequencies ideally suited for
measurement by LISA.  These waves come from the small body spiraling
through the black hole's very strong field --- regions very close to
its event horizon.  The number of orbits executed as the small body
spirals inwards is very large --- it orbits about $10^5 - 10^6$ times
before reaching a dynamical instability and plunging into the hole
(depending on the specific details of the system).  By tracking the
gravitational wave's phase evolution over these many orbits, we will
be able to ``weigh'' the spacetime's multipoles.  Rotating black holes
have a family of multipole moments set entirely by the black hole's
mass and spin:
\begin{equation}
M_l + i S_l = M(i a)^l\;.
\label{eq:kerr_multipoles}
\end{equation}
In this equation, $l$ is a spherical-harmonic-like index, $M_l$ is the
$l$-th ``mass moment'', and $S_l$ is the $l$-th ``current moment''.
(The index $m$ does not enter because quiescent black holes are
axisymmetric: all multipoles with $m \ne 0$ are zero.)  For a fluid
body, with density $\rho(\vec r)$ and velocity distribution ${\vec
v}(\vec r)$, the moments would be
\begin{equation}
M_l \sim \int d^3r\, r^l \rho({\vec r})\;,\qquad
S_l \sim \int d^3r\, r^{(l - l)} |{\vec r}\times{\vec v}(\vec r)|
\rho({\vec r})\;.
\label{eq:fluid_multipoles}
\end{equation}
A rotating black hole has a family of mass multipoles that is purely
even, a family of current multipoles that is purely odd, and is
determined entirely by the first two moments: $M_0 = M$ and $S_1 = a M
= |{\vec S}|$.  That these two moments --- the black hole's mass and
spin --- determine the {\it entire} family of multipole moments is a
restatement of the no-hair theorem: {\it A black hole's properties are
entirely determined by its mass and spin}.  (We ignore charged black
holes, which are of little astrophysical relevance.)  Measuring three
multipoles is sufficient to falsify whether the ``large central
object'' is in fact an Einstein black hole: if the third moment is
inconsistent with the first two according to Eq.\
(\ref{eq:kerr_multipoles}) then that object {\it is not} a black hole
as described by Einstein's theory of gravity.  On the other hand, if
all multipoles that we measure beyond the first two are consistent,
this is very strong evidence that the ``large central object'' is in
fact described by the Kerr black hole solution.

Mapping a planet's gravitational multipoles from satellite orbits is a
well-established science called {\it geodesy}.  For example, consider
a planet that is mostly spherical, with a small quadrupolar
distortion.  Orbits of this body will precess: the time to cover its
full range of $\theta$ motion is incommensurate with the time it takes
to move through the full range of $\phi$.  Measuring this precession,
we measure the quadrupolar distortion.  Mapping the multipoles of a
black hole's spacetime with small body orbits is very similar; in
recognition of this similarity, it has been suggested that the mapping
of black hole spacetimes be called {\it bothrodesy}
{\cite{bothrodesy}}.

A proof-of-principle analysis by Ryan {\cite{fintanmeasure}} has shown
that bothrodesy can be done with gravitational waves, showing that a
massive body's multipole moments are encoded in the waves emitted as a
small body spirals in.  His analysis is in somewhat crude, however.
He restricts the inspiraling body to circular, equatorial orbits,
losing information about the body's multipoles that are encoded in an
inclined orbit's precession.  Instead, he ``weighs'' the multipoles by
the fact that their radial dependence is different, and so affect the
orbit at different rates as the body spirals inward.  Even with this
excessively restricted setup, Ryan finds that at least three and in
some cases four or five multipoles can be distinguished in
gravitational-wave data.  It is almost certain we will find that
multipoles are measured even better when realistic orbits are used to
generate the waves.

\begin{figure}[t]
\includegraphics[width = 8.4cm]{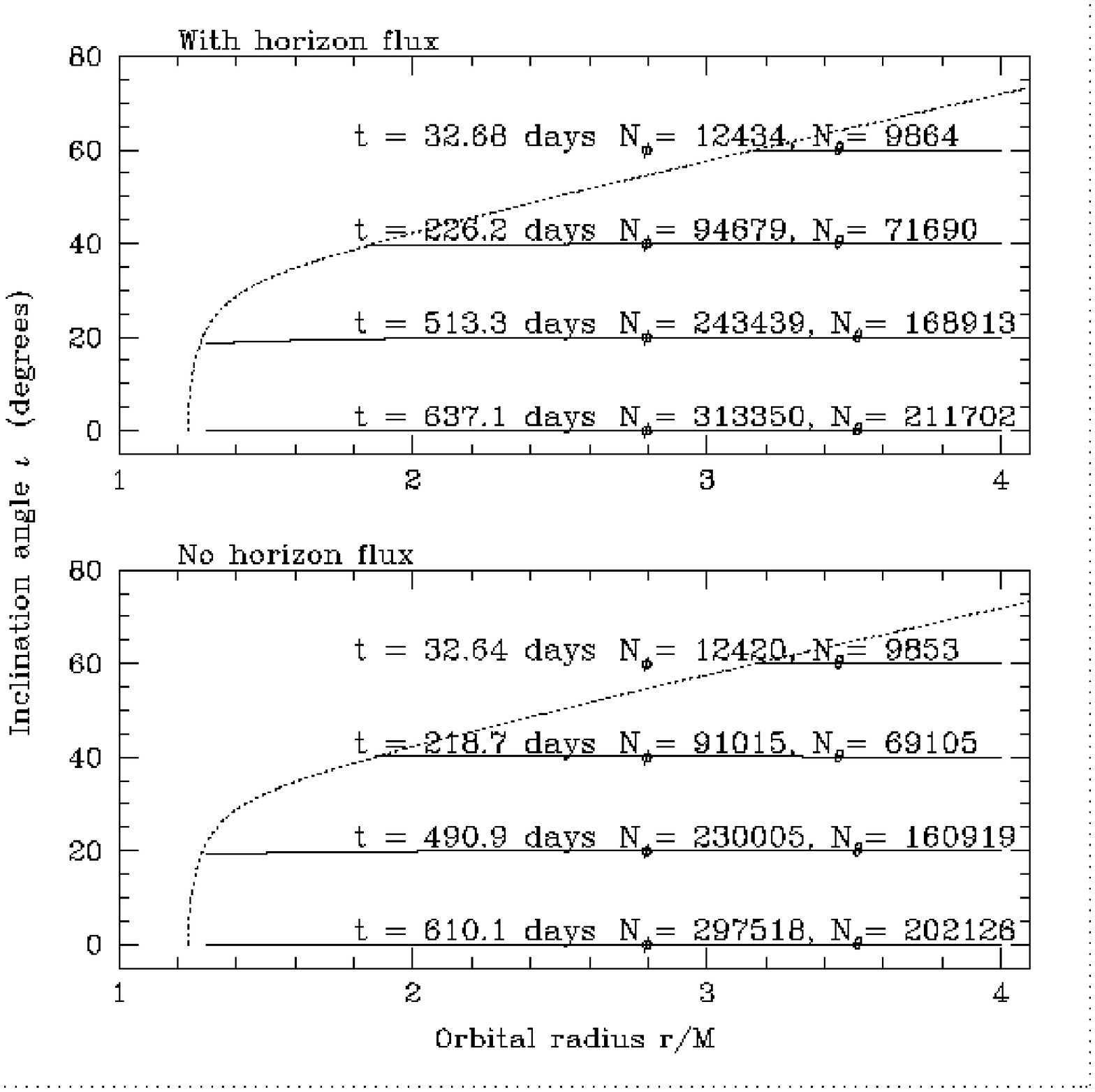}
\includegraphics[width = 8.4cm]{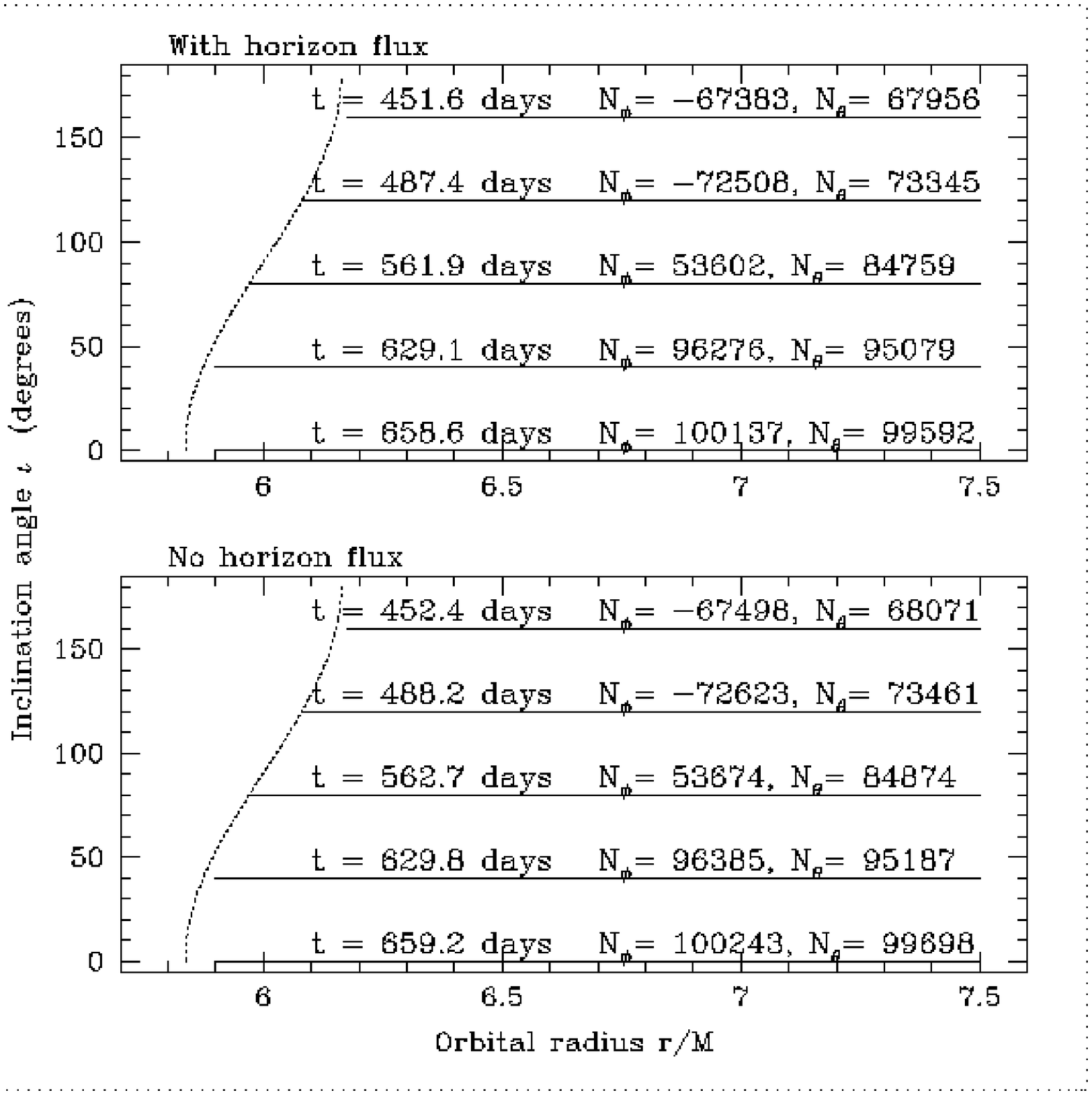}
\caption{Inspiral trajectories for small bodies spiralling into massive,
rotating black hole holes.  These trajectories assume that the large
black hole has mass $M = 10^6\,M_\odot$ and that the small body is
$1\,M_\odot$.  The trajectories on the left are for spiral into a
black hole with spin $a = 0.998M$ (a near maximally spinning hole);
those on the right are a hole with $a = 0.05M$.  Each point represents
a possible circular orbit, with radius $r$ and inclination to the
black hole's equator $\iota$.  Dotted lines show the innermost stable
circular orbit: orbits below and to the right of the dotted line are
stable, while those above and to the left are unstable, tending to
plunge into the black hole.  Gravitational waves drive the small body
to spiral through a sequence of circular orbits.  We show the time it
takes the body to spiral to the innermost stable orbit, the number of
orbits, $N_\phi$, executed as it does so, and the accumulated number
of oscillations in $\theta$, $N_\theta$.  In each figure, the top and
bottom panels compare inspiral with the influence of the hole's event
horizon included (top) versus ignored (bottom).}
\label{fig:inspiral_trajs}
\end{figure}

The black hole nature of a massive object influences the inspiral of a
small body in other ways.  Fig.\ {\ref{fig:inspiral_trajs}} shows the
``trajectories'' in radius $r$ and inclination angle $\iota$ followed
by a $1\,M_\odot$ body that spirals into a $10^6\,M_\odot$ black hole.
These trajectories were computed by studying how the energy and
angular momentum of the orbit evolve due to losses from gravitational
wave emission; see {\cite{hughes_circI,hughes_circII}} for details.
In the top panels, gravitational waves carry energy and angular
momentum both out to infinity and down the hole's event horizon.  In
the lower panels, the horizon flux is ignored.  When the hole is
rapidly spinning, the horizon flux has an enormous influence on the
inspiral, increasing the inspiral by several weeks and adding
thousands of additional orbits.  That the horizon flux {\it increases}
the inspiral duration is at first sight counterintuitive --- one would
guess that the horizon is a sink for the orbit's energy, and so its
influence should decrease the inspiral duration.  In fact, the hole
{\it supplies} orbital energy: if it were not for the loss of energy
in radiation carried to infinity, the body would spiral outward,
rather than inward!  This surprising behavior can be understood as
either a tidal interaction between the small body and the event
horizon, or as due to radiation that superradiantly scatters from the
ergosphere of the black hole; see Refs.\
{\cite{hughes_circII,teuk_press,hartle}} for further discussion.  When
the hole does not rotate rapidly, the effect is far smaller, and is
more intuitively acceptable: orbital energy is lost to the black hole.
Rapidly rotating black holes, if supplied to us by nature, will be
particularly interesting laboratories for probing strong field
gravity.

\subsection{Dense nuclear matter and gravitational waves}
\label{subsec:dense_nuclear}

Black holes are purely vacuum solutions to Einstein's equations.
Other gravitational-wave sources contain dense matter concentrations.
In particular, neutron star sources are among the most studied for
gravitational-wave detectors.  A typical neutron star consists of
approximately $1.5\,M_\odot$ of material with mean density $\rho \sim
10^{14}$ grams per cubic centimeter; such a star has a radius of 10 --
15 kilometers.  The properties of that matter can have a large
influence on the gravitational waves that neutron star systems
generate.  Thus gravitational-wave measurements from such systems may
be used to study highly dense matter.  Just a few measurements are
needed to strongly constrain the equation of state of neutron star
matter {\cite{lindblom_eos}}.  In this subsection, we briefly describe
some examples of how this may be done.

When a neutron star is a member of a binary, the details of its
nuclear matter structure are entirely irrelevant to gravitational-wave
generation over much of the inspiral.  It is only when the members of
the binary are close enough that their tidal gravitational fields
begin to distort one another that neutron star structure becomes
potentially important.  For double neutron star binaries, this means
that the very late inspiral and merger waves (when the neutron stars
collide and fuse into a single body) will depend quite strongly upon
the properties of dense nuclear matter.  Unfortunately, these waves
are at very high frequencies ($\sim1000$ Hz) where gravitational-wave
detectors do not have good sensitivity.  Special high frequency laser
interferometers {\cite{gdsw,meers,vmmb,klm,mizunoetal}} and acoustic
detectors {\cite{merk_johnson,hsp}} can be used to improve our ability
to measure these waves {\cite{last3min,hughes_thesis}}.

Better accuracy may be achieved by measuring waves from the
coalescence of a black hole-neutron star binary.  If the black hole is
more massive than roughly $10\,M_\odot$, the neutron star will
essentially be swallowed whole by the black hole.  This is because the
radii of the innermost orbits scale with the hole's mass, and because
the tidal field of the black hole scales as $M/r^3\sim 1/M^2$.  Hence
when the hole is large, its tides are relatively gentle and the
neutron star can survive inspiral all the way through the hole's event
horizon.  When the hole is smaller, however, the tides rip the neutron
star apart before any matter falls down the horizon.  This drastically
changes the binary's mass distribution and significantly affects its
gravitational-wave emission.  Because a black hole-neutron star system
is significantly more massive than a double neutron star system, the
waves that depend upon the neutron star's matter are emitted at lower
frequencies where laser interferometers have better sensitivity.
Vallisneri {\cite{vallis}} has shown that planned upgrades to LIGO
will make it possible to extract a fair amount of information about
neutron star properties by measuring such disruptions.  However, he
cautions that better theoretical modeling of the waves'
characteristics will be needed to fully understand what the detectors
measure.

Neutron stars may undergo phase transitions as dense material changes
state, perhaps forming a condensation of strange matter or free quark
matter {\cite{glendenning}}.  If this occurs, the star's density
profile will rapidly change; any gravitational waves that it might be
emitting will rapidly change character as well.  This may be seen by
following the waves that are emitted by an accreting neutron star
{\cite{chengdai}}: matter dumped onto the neutron star increases its
mass beyond the critical point, until the phase transition occurs.
The star then changes radius rapidly, exciting radial vibrational
modes in the star.  Other work suggests that such a phase transition
may be see in the waves that are emitted as a neutron star forms
during supernova collapse {\cite{stm}}, changing the character of the
star's various oscillational modes.  Finally, we note that a
transition to such a form of nuclear matter has a drastic effect on
the viscosity of material by changing the kinds of interactions that
occur between nucleons in the star.  This may cause various
instabilities, particularly that due to r-modes, to damp out rapidly
{\cite{jones_prd,jones_prl,owen_lindblom}}.  Gravitational-wave
generation by neutron star systems present great opportunities for
studying extremely dense matter, but as Vallisneri has emphasized,
many theoretical uncertainties must be hammered out before we will be
able to do so well.

\subsection{Tests of new physics}
\label{subsec:new_physics}

Finally, gravitational waves will test current ideas for new
unification physics beyond the standard-model-plus-GR paradigm.  The
best models for fundamental unification, based on string theory, lead
to the idea that gravity propagates in more than three spatial
dimensions.  This has led to the investigation of ``brane worlds'', in
which standard model fields live on a 3D ``brane'' or boundary of a
larger space (``bulk'') with one or more small or highly-curved extra
dimensions
{\cite{akama,holdom,rub_shap,add,aadd,randallsundrum1,randallsundrum2}}.
In these worlds, gravitons can have massive ``Kaluza-Klein'' modes
that might be produced in accelerators, or survive from the early
universe as dark matter. Even the massless modes, classical
gravitational waves, no longer necessarily obey the same causal
structure as electromagnetic waves; they can propagate either above or
below the speed of light
{\cite{chung_freese,ishihara,ceg,hebecker_marchrussell,moffatt,youm,caldwelllanglois}}.

The possibility that a graviton might travel slower than a photon
appears to be very tightly constrained by the fact that ultra high
energy cosmic rays do not appear to lose energy to gravitational {\v
Cerenkov} radiation {\cite{moorenelson}}.  However, in brane worlds,
gravitons could also propagate faster than the speed of light.  One
can think of the extra dimensions as a kind of waveguide for the
gravitational waves, so that the propagation is nondispersive;
however, the curvature of the extra dimensions can also allow for
faster propagation. We have little empirical information about this
possibility at present, since we know only that the right amount of
gravitational wave power is being emitted from binary pulsars. With
LISA we will have coherent streams of gravitational waves from
identified compact white-dwarf binaries that also have optically
measured orbital parameters and orbital phases. The optical and
gravitational wave data can be compared over many orbits, yielding a
direct test of the propagation speed of gravitational waves that
should be very precise.

These data will also constrain the possibility that the graviton is a
composite particle, with components of more than one (possibly
nonzero, but very small) mass. Classically emitted waves may, after
propagating a large distance, separate out into two or more
components. We might see a ``standard'' waveform of reduced amplitude,
corresponding to the massless mode, with the rest of the energy
appearing in some other, more dispersive waveform (or perhaps so
highly delayed that it is not seen at all).

While we have discussed these programs in the context of concrete
brane world models, it is possible to motivate them in more general
language, as tests of Lorentz symmetry
{\cite{peters,colemanglashow,steckerglashow}}.  The proliferation of
unification ideas certainly justifies such an empirical approach; the
detection of a cosmological constant has opened up consideration of an
enormous range of mass scales so it has become interesting to
constrain even what used to seem outrageously small masses.

Cosmology provides another context in which gravitational wave
detectors will constrain fundamental physics.  Generally, the
connection comes through the possibility of generating a stochastic
background of gravitational waves in the early universe that can be
detected today {\cite{maggiore}}.  Gravitational-wave observatories
will have an interesting sensitivity to these backgrounds; both LIGO
and LISA will be able to detect broad-band stochastic backgrounds with
energy densities well below the microwave background
{\cite{hoganbender,bruceleshouches}}.

The source most often considered for such a background is cosmological
inflation. We now think that parametrically-amplified quantum
fluctuations of the inflaton field are responsible for large scale
structure in the universe, and also for most of the fluctuation power
in the microwave background anisotropy. The model predicts that
similar fluctuations in the graviton field will produce a
gravitational wave background. It is possible that these tensor waves
are already contributing to the detected microwave anisotropy on the
largest scales, and indeed there are experiments to test this
possibility using the different polarization signatures of the
anisotropy produced by the two kinds of modes.

These experiments are clearly important, but it will not be surprising
if the tensor component turns out to be negligible. There is no reason
for it to be comparable to the scalar (inflaton produced) component,
as it depends on different parameters of the inflation model. (Indeed,
the pure-scalar model with a scale-invariant spectrum fits the
anisotropy data so well we would be disappointed to spoil it by adding
another component with extra parameters.)  The likely outcome is that
the polarization experiments will constrain the tensor component,
thereby constraining directly the expansion rate during inflation.

Even if the tensor component is detectable on the largest scales, the
corresponding gravitational wave energy density for a scale-free
spectrum (about $10^{-10}$ of the critical density at low frequencies,
changing to $10^{-10}$ of the microwave energy density at high
frequencies) is far too low for LIGO or LISA to detect directly. In
some models, where the expansion rate increased to nearly the Planck
scale during the course of inflation, the spectrum can be tilted
enough to produce a detectable background, but these models are
already ruled out by millisecond pulsar timing (which limits
stochastic backgrounds to less than $10^{-3}$ of the microwave
background density at frequencies of the order of inverse years)
{\cite{kaspi_taylor_ryba,thorsett_dewey}}.  Therefore it seems
unlikely that the quantum graviton background will be seen by these
observatories.

Some models however predict an entirely different source of primordial
backgrounds. After inflation, the early universe may have hosted
violent classical events that gave rise to a stochastic
background. These gravitational waves could be much more intense than
the inflationary quantum graviton production; the maximal energy
density, limited by equipartition arguments, is approximately the same
as the microwave background energy density--- ten orders of magnitude
above the typical inflationary level.

There are a number of circumstances which might approach within a few
orders of magnitude of the maximal level. For example, a strongly
first-order phase transition would lead to bubble nucleation and
relativistic flows on the Hubble scale
{\cite{witten,hogan_mnras,kkt}}; condensation of brane in a
higher-dimensional space would lead to Kibble excitation of tensor
modes on the Hubble scale {\cite{hogan_prl,hogan_prd}}.  Strong
effects of this kind are not required in any well established theory,
but they are generic behaviors of relativistic systems that are worth
constraining.

Backgrounds from these events are not scale-free; they tend to be
broad band, but they have a characteristic frequency given by the
redshifted Hubble rate. For detection at LISA and LIGO frequencies,
the corresponding cosmic temperature turns out to be upwards of TeV
energies--- a range of interest for unification and supersymmetry
breaking. This is a period of cosmic history about which we have
almost no empirical information at present, so even upper limits from
gravitational wave observatories will convey new information about
physics at these energies. A detection would open up a new period of
cosmic history to detailed scrutiny; we might for example find much
more detailed information about the process of cosmic
baryogenesis. The ``last scattering surface'' for gravitational waves
is just their production epoch (that is, there is almost no
significant absorption or large-angle scattering since the Planck
time) so they provide a direct window into the early universe much
deeper than the microwave background, and down to much smaller
comoving scales.  If a detectable stochastic background does exist,
future generations of detectors may one day measure anisotropy in the
background that contains new information about the early universe ---
a directly-imaged view of events invisible (even in principle) in any
other way.

\begin{acknowledgments}

SAH is supported by NSF Grant PHY--9907949; his travel to the Snowmass
meeting was in addition supported in part by the LISA Project and the
Jet Propulsion Laboratory, California Institute of Technology under
contract to NASA.  LIGO is funded by the National Science Foundation
under Cooperative Agreement PHY--9210038.  This work was supported at
the University of Washington by NSF Grant AST--0098557, and at the
University of Colorado by NASA Grants NAG5--4095 and NAG5--10259.

We would like to thank colleagues who helped us and gave excellent
talks at Snowmass 2001 P4.6: Keith Riles, Nergis Mavalvala, Rana
Adhikari of LIGO.  David Shoemaker kindly provided us with excellent
LIGO posters and support of our efforts, Robin Stebbins provided much
of the LISA graphics used here, and Kip Thorne provided us with the
figure of LIGO sources.  We thank Yuk Tung Liu for helpful comments on
gravitational waves from accretion induced collapse of white dwarfs.
We appreciate the support, help and encouragement of Barry Barish,
Stan Whitcomb, and Tom Prince.

\end{acknowledgments}

%
%


\begin{thebibliography}{99}
\bibitem{explain_sensitivity} Gravitational-wave strain sensitivies
	are usually quoted with units ${\rm Hz}^{-1/2}$.  This is
	because the squared strain noise in a band $\Delta f$ is given
	by $\sigma_h^2 = \int_{\Delta f} df\,{\tilde h}^2$; the final
	strain noise is thus dimensionless.  The noise spectrum
	${\tilde h}^2$ is often denoted $S_h$.
\bibitem{bothrodesy} The term ``bothrodesy'' has been coined by
	Sterl Phinney.  It descends from the use of the suffix
	``-bothron'' introduced by Brandon Carter to describe
	properties of black holes.  This suffix, in turn, descends
	from the Greek word ``$\beta o\theta{\kern -0.09em}\rho
	o\varsigma$'', meaning ``garbage dump''.
\end{thebibliography}
\end{document}